\documentclass{aa}
\usepackage{graphicx}
\usepackage{hyperref}
\usepackage{txfonts}
\usepackage{xcolor}
\usepackage{caption}
\usepackage{subcaption}
\usepackage{textcomp}
\usepackage[rightcaption]{sidecap}
\usepackage[desactivate]{linenoaa}
\usepackage{placeins}  

\graphicspath{{Figures/}}

\newcommand{\CellWithForceBreak}[2][c]{
\begin{tabular}[#1]{@{}c@{}}#2\end{tabular}}

\begin{document} 
    \title{Coronal dimmings from active region 13664 during the May 2024 solar energetic events}
    \titlerunning{Coronal dimmings from active region 13664 during the May 2024 solar energetic events}	
	
    \author{Amaia Razquin
          \inst{1}
          \and
          Astrid M. Veronig\inst{1, 2}
          \and
          Karin Dissauer\inst{3}
          \and 
          Tatiana Podladchikova\inst{4}
          \and 
          Shantanu Jain\inst{4}
    }

    \institute{University of Graz, Institute of Physics, Universitätsplatz 5, 8010 Graz, Austria \\
        \email{amaia.razquin-lizarraga@uni-graz.at}
        \and 
            University of Graz, Kanzelh\"ohe Observatory for Solar and Environmental Research, Kanzelh\"ohe 19, 9521 Treffen, Austria
        \and
             NorthWest Research Associates, 3380 Mitchell Lane, Boulder, CO 80301, USA
        \and
            Skolkovo Institute of Science and Technology, Bolshoy Boulevard 30, bld. 1, 121205 Moscow, Russia
    }
	
    \date{Received 26 March 2025; accepted 13 May 2025}
	
    \abstract 
    {Coronal dimmings are regions of transiently reduced brightness in extreme ultraviolet (EUV) and soft X-ray (SXR) emissions associated with coronal mass ejections (CMEs), providing key insights into CME initiation and early evolution. During May 2024, AR~13664 was among the most flare-productive regions in recent decades, generating 55 M-class and 12 X-class flares along with multiple Earth-directed CMEs. The rapid succession of these CMEs triggered the most intense geomagnetic storm in two decades.
    }
    {We study coronal dimmings from a single active region (AR~13664) and compare them with statistical dimming properties. We investigate how coronal dimming parameters -- such as area, brightness, and magnetic flux -- relate to key flare and CME properties.
    }  
    {We performed coronal dimming detection on observations from the Atmospheric Imaging Assembly (AIA) instrument on board the Solar Dynamics Observatory (SDO). We used a logarithmic base-ratio thresholding technique to identify dimming regions, selecting pixels where $\log{(I/I_0)}\leq -0.19$. Due to the high activity of the AR, we propose a quantitative threshold for distinguishing real mass depletion dimmings from unrelated intensity reductions by setting a threshold on the dimming area reached within the first hour ($A\geq6.48\times10^9$~km$^2$). We systematically identified all flares $\geq$M1.0, all coronal dimmings and all CMEs (from the CDAW SOHO/LASCO catalogue) produced by AR~13664 during 2024 May 1 - 15, and studied the associations between the different phenomena and their characteristic parameters.
    }
    {We detect coronal dimmings in 22 events, with 16 occurring on-disc and six off-limb. Approximately 83\% of X-class flares and 23\% of M-class flares are associated with CMEs, with 13 out of 16 on-disc dimmings linked to CME activity. The dimmings in AR~13664 exhibit total unsigned magnetic fluxes exceeding  $5.5\times 10^{21}$~Mx, reflecting the region's high magnetic flux density; and dimming areas greater than $1.16\times10^{10}$~km$^2$. 
    Previous statistical studies had shown a correlation between dimming parameters and flare parameters. 
    We find that dimming parameters for the May 2024 events, particularly total dimming area and area growth rate, have a stronger correlation with GOES soft X-ray peak flux and fluence than anticipated, highlighting the connection between energy release in flares and the accompanying dimming. We find correlations between dimming properties and CME maximum velocities, which indicate that coronal dimmings serve as proxies for CME speeds. 
    }
    {Our results support the strong interplay between coronal dimmings and flares, as we find increased correlations between flare and dimming parameters in this single-AR study compared to the general dimming population. Furthermore, we confirm that coronagraphic observations, unable to observe the lower corona, underestimate correlations between CME velocities and dimming parameters, as they fail to capture the early CME acceleration phase. This highlights the critical role of dimming observations in providing a more comprehensive understanding of CME dynamics.}
    \keywords{Sun  --
		dimmings  --
		solar activity --
		coronal mass ejections -- 
        flares --
		May 2024 storms
    }
	
    \authorrunning{A. Razquin et al.}
    \maketitle

\section{Introduction}
Coronal dimmings are transient reductions in extreme ultraviolet (EUV) and soft X-ray (SXR) emissions from the solar corona, closely associated with coronal mass ejections (CMEs) \citep{hudson1996long, sterling1997yohkoh, thompson1998soho, thomson2000coronal}.
These dimmings arise from a sudden density depletion caused by the evacuation of plasma during the early CME evolution as interpreted from multiwavelength observations \citep{Zarro1999soho}, as well as spectroscopic \citep{harra2001material, Jin2009coronal, Veronig2019} and differential emission measure \citep{vanninathan2018plasma} analyses. Consequently, coronal dimmings serve as key indicators of plasma outflows resulting from the initial expansion of CME structures and overlying fields.

Earth-directed CMEs are the primary drivers of severe space weather events that impact the near-Earth environment (e.g. \citet{gopalswamy2009soho}). They are traditionally observed using whitelight coronagraphs and in situ measurements; however, these methods do not provide accurate constraints on their early evolution and characteristic properties \citep{burkepile2004role}. On the solar disc, localised dimmings with a pronounced decrease in brightness -- referred to as core dimmings -- are typically rooted in the different polarities of the footpoints of the erupting flux rope \citep{thompson1998soho, attrill2006using}, while extended and shallow dimmings -- secondary dimmings -- show the density depletion due to the overlying fields stretching and partially reconnecting \citep{mandrini2007cme, Dissauer2018b}. Coronal dimmings have been thus studied to get valuable insights into CME onset, early evolution, and plasma characteristics (e.g. \citet{james2017ondisc}; \citet{Veronig2019}; \citet{Podladchikova2024}). In addition, they give also important insights into the magnetic flux systems involved in the eruption (see recent review by \citet{Veronig2025}). 

Statistical studies have established correlations between on-disc coronal dimmings and key CME properties such as mass, acceleration, and speed, as well as the characteristics of associated flares \citep{reinard2009relationship, aschwanden2016global, krista2017statistical, Dissauer2018b, Dissauer2019}. Similarly, off-limb coronal dimmings have also been statistically linked to CMEs \citep{bewsher2008relationship, aschwanden2017global, chikunova2020coronal}. Furthermore, dimmings have been detected in spatially integrated data \citep{mason2016relationship}, showing a strong connection to CME mass and speed, and have been consequently proposed as a potential proxy for stellar CMEs \citep{Jin2020coronal, Veronig2021indications}. Additionally, recent studies suggest that dimming expansion provides additional insights into the CME expansion and recovery processes \citep{ronca2024recovery} and can also be used to estimate their initial propagation direction \citep{Chikunova2023, jain2024coronal, Podladchikova2024}. A recent example demonstrated the application of the DIRECD method to predict the CME's propagation direction for a halo CME on May 8, underscoring the potential of dimming analysis for early assessment of CME evolution \citep{jain2024estimating}.

In May 2024, NOAA active region (AR) 13664 was the source of the largest geomagnetic storm since 2003 \citep{hayakawa2024solar}. AR~13664  was visible from Earth for the first time on the south-eastern limb on May 1 as a bipolar sunspot group. From May 2 to 7, it rapidly grew in size from 110 to 2700 $\mu$hem, reaching a size comparable to the AR that originated the Carrington event \citep{carrington1859description}. 
AR~13664 developed into a Fkc McIntosh type, and from May 6 it was of magnetic class $\beta\gamma\sigma$ \citep{hayakawa2024solar}. Starting from May 7, the magnetic flux and free magnetic energy of AR~13664 was strongly increasing \citep{Jarolim2024magnetic, hayakawa2024solar}. 

\citet{hayakawa2024solar} gave a first overview on the solar events and storms produced by AR~13664. They report that it produced 14 major CMEs between May 8 and May 14, of which 10 were halo CMEs. The rapid succession of CMEs caused the strongest geomagnetic storm since the Halloween events in November 2003, with a Dst peak value of $-412$~nT on May 11 around 4~UT. During the geomagnetic storm, auroras were reported as far south as 21$^\circ$~N latitude and $30^\circ$~S over the Western African and Atlantic sectors \citep{Karan2024GOLD}. It caused increased ionospheric activity with a significant depression of the ionospheric plasma \citep{spogli2024effects}, and more than 5,000 spacecraft had to perform altitude changes to counteract the enhanced orbital decay \citep{parker2024satellite}.

\section{Data and event identification} \label{sec:data}
In this paper, we perform a comprehensive analysis of coronal dimmings and associated flares and CMEs that originated in AR~13664 over its first disc passage from May 1 to 15, 2024. To this aim, we start with identifying the major flares produced by the AR (defined as flares of GOES class M1.0 and larger), and then check for CMEs and coronal dimmings that are associated with these flares.

\subsection{Event selection and flare-CME-dimming association} \label{sec:data:selection}
For the identification of the GOES X-ray flares, we used the NOAA Solar and Geophysical Event Reports provided by the Space Weather Prediction Center \footnote{\url{https://www.star.nesdis.noaa.gov/goes/index.php}}. In two cases where flares occurred in close succession (with start times within 30~min from each other), we treated them as a single event, characterising it based on the maximum GOES SXR flux and defining its duration from the start of the first flare to the end of the last. The first of these two events is a combination of an M3.1 and an M2.9 flare on May 9 at 11:52~UT and 12:05~UT, respectively. The second combination involves three flares on May 13, starting at 08:06~UT (M1.2) 08:23~UT (M1.4) and 08:48~UT (M6.6). This gives a total of 67 flares, comprising 12 X-class flares and 55 M-class flares.

The CMEs from  AR~13664 were identified from the measurements of the LASCO C2 and C3 coronagraphs compiled in the CDAW SOHO/LASCO CME catalogue\footnote{\url{https://cdaw.gsfc.nasa.gov/CME_list/}} as described in \citet{yashiro2004catalog}. We applied a systematic approach to associate CMEs with flares originating from AR~13664 based on their timing and spatial proximity. Firstly, we identified CMEs initially detected within two hours of a flare's onset and paired them with the corresponding flare. Secondly, only pairs where the flare location was within the total angular extent of the CME were kept. Finally, we conducted a visual inspection to eliminate mismatched flare-CME pairs. The list of CMEs was then cross-referenced with the list given by \citet{hayakawa2024solar}. To extract the CME velocity, we used the height-time profiles provided in the catalogue. We associated 23 CMEs to AR~13664, ten of which were linked to X-class flares and 13 to M-class flares. 

We analysed EUV images for coronal dimmings for each individual flare event, establishing a default flare–dimming association. We identified 22 coronal dimmings, with ten linked to X-class flares and 12 to M-class flares. Additionally, since CMEs were matched to flares as previously described, we also established a dimming–CME connection. However, one case presented an exception: the M1.9 flare at 03:19~UT on May 8 was accompanied by a dimming but had no associated CME. Instead, a CME was linked to an M3.5 flare that occurred roughly an hour earlier but was not accompanied by a dimming. Given the timing and location of the events, we chose to associate this CME also to the dimming observed in the later flare. Consequently, 19 dimming detections were associated with CMEs, which left four CMEs without associated dimming and three dimmings without associated CME. 

The identified M- and X-class flares, associated CMEs and coronal dimmings are listed in Table~\ref{table:flare_list}. The two events containing multiple flares are shown with the distinct timing of each flare, but with the same identification number (`N', first column) and combined flare properties. The associated CMEs are identified with the ID used in the CDAW catalogue, which represents the day and time of the first appearance in the LASCO C2 field of view. The CMEs with an asterisk (*) are not present in the list by \citet{hayakawa2024solar}. 

\begin{figure}[t] 
    \resizebox{\hsize}{!}{\includegraphics{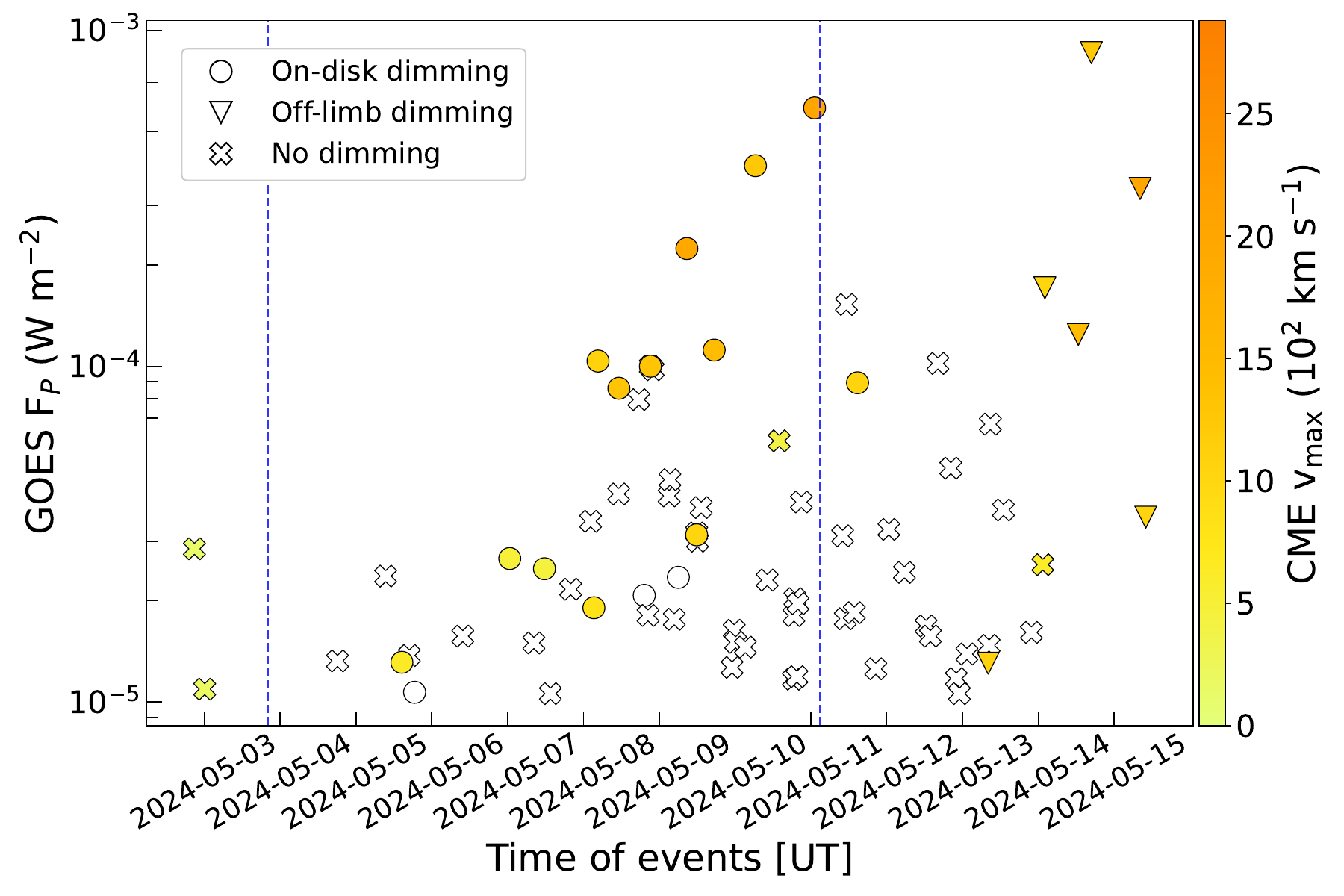}}
	\caption{GOES SXR peak flux of M- and X-class flares from AR~13664 between May 1 and May 15, 2024. Flares with an associated CME are coloured based on the maximum speed of the CME (see colour bar to the right), flares with no CME are indicated by open markers. Circles show flares with associated on-disc dimmings, triangles indicate flares with off-limb dimmings, and crosses indicate flares with no associated dimming. Vertical dashed blue lines mark the period when AR~13664 is within $45^\circ$ of the disc centre, where dimming detection is more reliable.}
    \label{fig:flare_cme_dimming}
\end{figure}

Figure~\ref{fig:flare_cme_dimming} presents the GOES SXR peak flux of M- and X-class flares from AR~13664 observed between May 1 and May 15, 2024. Flares with associated CMEs are shown with circles colour-coded based on the maximum velocity of the CME (from the CDAW catalogue), while confined flares are indicated by open markers. The association with coronal dimmings is indicated with different markers: circles indicate flares with detected on-disc dimmings, triangles show flares with off-limb dimmings and crosses show flares with no associated dimming. The two vertical dashed blue lines highlight the time interval during which the AR was within $45^\circ$ of the disc centre, where dimming detection is more reliable. 

\subsection{Data and data reduction} \label{sec:data:reduction}

Coronal dimmings were examined through a sequence of full-cadence (12~s) Solar Dynamics Observatory (SDO; \citet{pesnell2012sdo})/Atmospheric Imaging Assembly (AIA; \citet{Lemen2012aia}) 211~\AA~images \texttt{aia.lev1\_euv\_12s}, which capture the AR corona at a temperature around $2\times 10^6$~K. The magnetic properties were analysed using the 720~s line-of-sight (LOS) magnetograms (\texttt{hmi.B\_720s}) from the SDO/Helioseismic and Magnetic Imager (HMI; \citet{Scherrer2012HMI, Schou2012design}). 

The data was rebinned in spatial resolution from $4096\times4096$ to $2048\times2048$ pixels under flux conservation conditions and was processed using standard Solarsoft IDL software (\texttt{aia\_prep.pro} and \texttt{hmi\_prep.pro}). For each flare event, the data was rotated to a common reference time using \texttt{drot\_map.pro} to account for differential rotation. The reference time was selected to be 15~min before the start of the flare. Only AIA images with an exposure time between 1.8 and 3.0~s were used for the analysis. The detection of dimmings was confined to a subfield of $1000\times 1000''$  located around the centre of the relevant flare. The evolution of the coronal dimming was examined for a duration of 2~hr 15~min, starting 15~min before the onset of the flare. 

However, for some events a subsequent flare occurred earlier than 2~hr; in such cases, the detection was manually stopped at the start of the subsequent flare. A single magnetogram recorded 15~min before the flare start was used for the analysis of magnetic properties. Several events, particularly those occurring after 17:00~UT on May 8, required an alternative data reduction algorithm owing to a gap in SDO data. This other processing is explained in Appendix~\ref{Appendix_alternative} and the flares that were analysed using this processing appear marked with a caret ($^\wedge$) on Table~\ref{table:flare_list}.

\section{Methods} \label{sec:methods}

We conducted an analysis of coronal dimmings of AR~13664 observed in the 211~\AA~channel of SDO/AIA and examined these events in the context of the statistical dimming dataset analysed by \citet{Dissauer2018b, Dissauer2019} by comparing key properties, validating and extending previous findings. In Sects.~\ref{sec:methods:detection} and \ref{sec:methods:parameters} we outline the dimming detection algorithm and provide a description of the characteristic parameters that define their physical properties based on \citet{Dissauer2018a}. We use the X1.1 event on May 9, associated with a halo CME, to illustrate the detection method and resulting evolution of the dimming parameters. Sect.~\ref{sec:methods:selection} describes the method used to discern false positive detections from real coronal dimmings employing the evolution of the dimming properties. In Sect.~\ref{sec:methods:statistical_analysis} we give an overview of the statistical analysis performed and the error calculation process.

\subsection{Dimming detection} \label{sec:methods:detection}

Dimming detection was performed on logarithmic base-ratio images using the region-growing algorithm described in \citet{Dissauer2018a}. The base image was built as the median in each pixel over the first three images within the time series which starts 15~min before the flare start time listed in Table~\ref{table:flare_list}. The pixels in subsequent images whose logarithmic (log10) ratio intensity decreased below a value of $-0.19$ were identified as dimming pixels. This intensity drop corresponds to about 35\% change in linear space. To reduce the noise a $3\times 3$ median filtering was applied and morphological operators were used to smooth the extracted regions and reduce the number of misidentified pixels. 
The detection in each image was saved in a cumulative dimming pixel mask where all pixels identified as dimming pixels up to the time $t_i$ were stored. From these cumulative dimming masks, timing maps were created, in which the time at which each pixel was detected as a dimming pixel for the first time was stored. In addition, minimum intensity maps were created by saving the minimum intensity of each pixel found in the final cumulative dimming mask during the considered time range $t_N$. The intensity values for logarithmic base-ratio images, base-difference images and direct images were calculated simultaneously. Besides on-disc dimmings, we also looked for off-limb dimmings using the same detection algorithm. In this paper, however, we focus the analysis on the characteristic parameters of on-disc dimmings only and report off-limb dimming detections for completeness.

\begin{figure*}[t] 
	\centering
        \includegraphics[width=17cm]{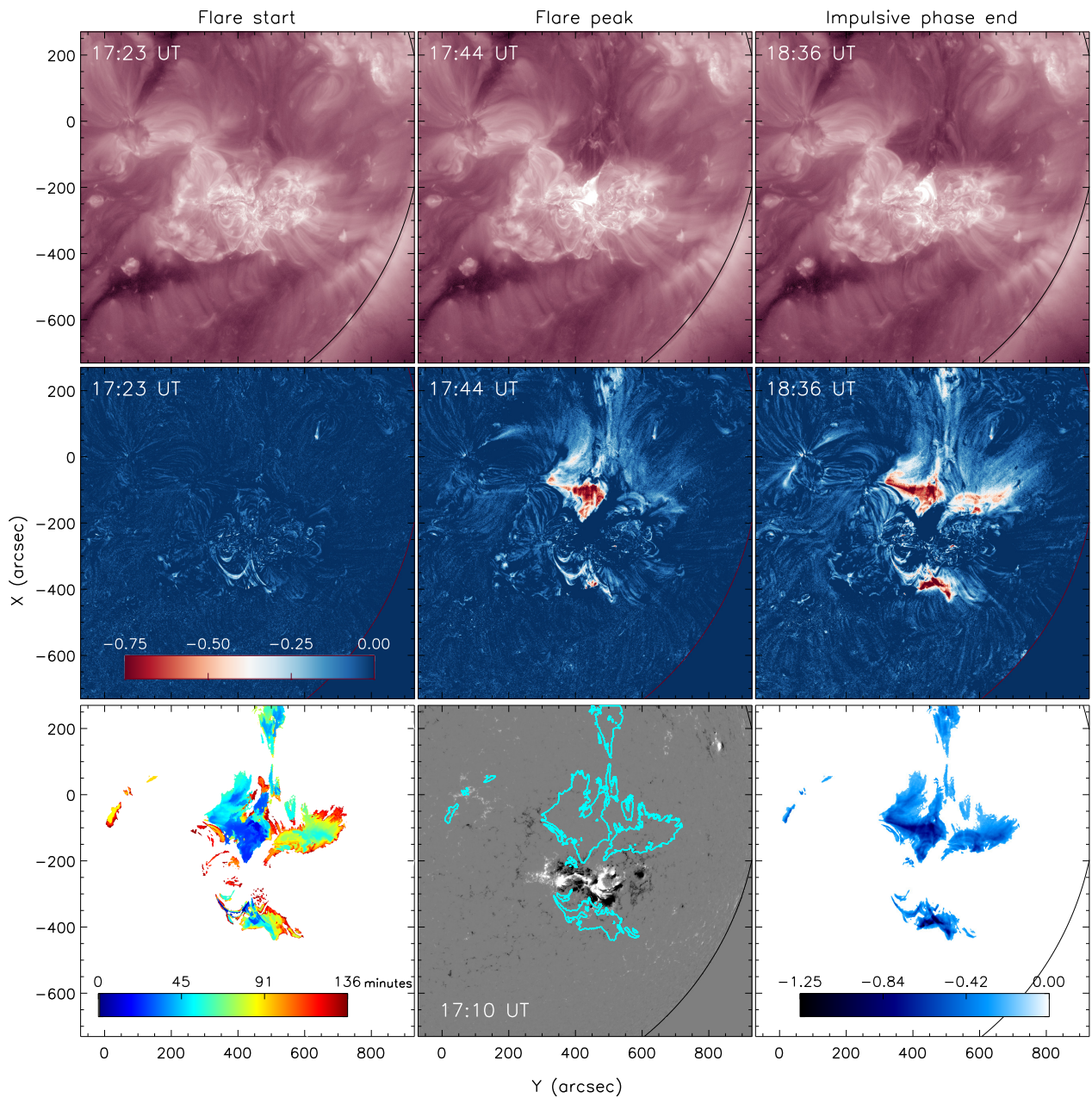}
	\caption{Overview of the evolution of the X1.1 flare and associated dimming on 2024 May 9 (no. 29). Top row: SDO/AIA 211~\AA~ direct images of AR~13664 at the flare's onset, peak time, and the end of the impulsive dimming phase. Second row: Corresponding logarithmic base-ratio images sharing the same colour scale, indicated by the bar in the left panel. Third row: Timing map indicating the first detection time of a dimming pixel (left), the HMI LOS magnetogram with the dimming region outlined in cyan contours (middle), and the minimum intensity map from logarithmic base-ratio data (right).}
	\label{fig:overview-example}
\end{figure*}

Figure~\ref{fig:overview-example} shows an overview of the dimming detection associated with flare no. 29, which was an X1.1 flare on May 9 (cf. Table~\ref{table:flare_list}), to illustrate the dimming detection method. The panels in the top row show SDO/AIA 211~\AA~images of AR~13664 at the flare start, flare peak and at the end of the impulsive dimming phase (see Sect.~\ref{sec:methods:parameters}). The panels in the middle row show the corresponding logarithmic base-ratio images. In these panels, the regions where the intensity decreases (the coronal dimmings) are shown from white to red, and regions where the intensity increases or remains unchanged are depicted in dark blue. The bottom left panel shows the timing map, in which each pixel is colour-coded according to the time of its first detection. The bottom middle panel shows the corresponding SDO/HMI LOS magnetogram, with the dimming region outlined in cyan contours. The bottom right panel shows the minimum intensity map recorded during the dimming impulsive phase in logarithmic base-ratio values.

\subsection{Dimming parameters} \label{sec:methods:parameters}
We derived characteristic parameters, such as dimming area, brightness, and magnetic flux, that describe the physical properties of coronal dimmings. The parameters selected for this study are a subset of those considered in \citet{Dissauer2018b, Dissauer2019}. The numerical values of the parameters are presented in Table~\ref{table:dimming_properties} for the flares in which an on-disc dimming was observed.

\begin{figure}[t] 
	\centering
    \resizebox{\hsize}{!}{\includegraphics{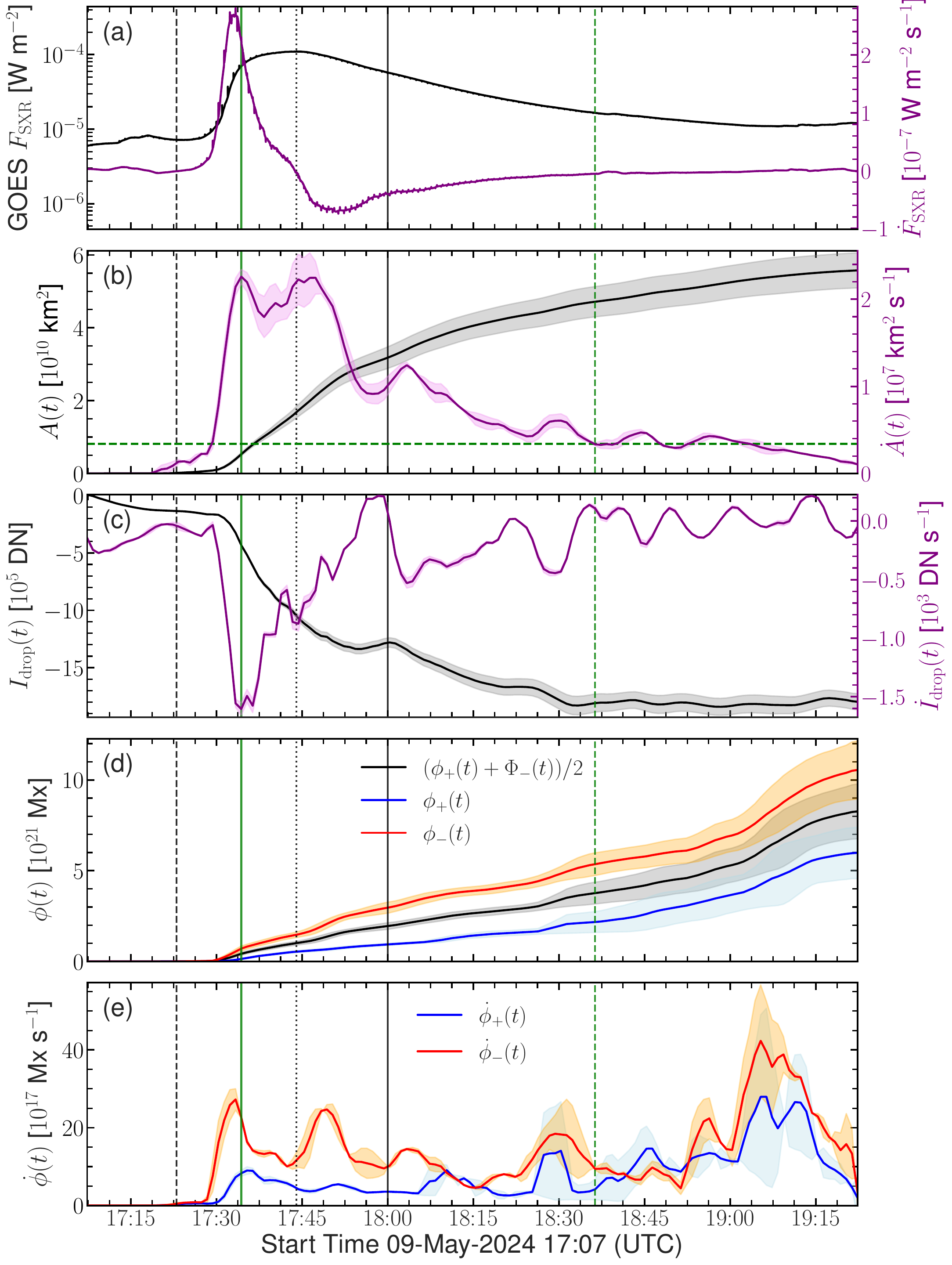}}
	\caption{Time evolution of selected dimming parameters for the X1.1 flare on 2024 May 9  (no. 29). From top to bottom: (a) GOES $1.0-8.0$~\AA~SXR flux (black) and its time derivative (purple); (b) expansion of the dimming area $A(t)$ (black) and the area growth rate, $\dot{A}(t)$ (purple), with the horizontal green dashed line marking when $\dot{A}(t)$ falls below 15\% of its maximum value which defines the end of the impulsive phase of the dimming; (c) brightness drop $I_\text{drop}(t)$ (black) and its change rate, $\dot{I}_\text{drop}(t)$ (purple) in base-difference units; (d) positive (blue), negative (red) and total unsigned magnetic flux (black) covered by the dimming regions; (e) corresponding magnetic flux rates, $\dot{\phi}_+(t)$ (blue), $|\dot{\phi}_-(t)|$ (red), and $\dot{\phi}$ (black).  Vertical dashed, dotted, and solid black lines indicate the start, peak, and end of the flare, respectively. The vertical solid and dashed green lines indicate the peak of $\dot{A}(t)$ and the end of the impulsive phase of the dimming, respectively.}
	\label{fig:evolution-example}
\end{figure}

The size of the dimming regions is represented by the dimming area, $A(t)$, which is defined as the cumulative area of all dimming pixels detected up to time $t$. From the time derivative of the dimming area, the area growth rate $\dot{A}(t)$ is calculated, which represents how fast the dimming region is growing. For event no. 29, Fig.~\ref{fig:evolution-example} (panels (b)-(e)) illustrates the evolution of the dimming parameters. The solid lines represent the mean values, while the shaded areas represent the $1\sigma$ standard deviation used as error bars for each parameter (see Sect.~\ref{sec:methods:statistical_analysis}). The evolution of the cumulative dimming area $A(t)$ is plotted in panel (b) in black and the area growth rate $\dot{A}(t)$ is shown in purple. The vertical solid green line represents the time at which $\dot{A}(t)$ reaches its maximum. Panel (a) shows the GOES SXR flux, $F_\text{SXR}$, (black) and its time derivative, $\dot{F}_\text{SXR}$, (purple).  The vertical dashed, dotted, and solid black lines indicate the start, peak, and end of the associated flare. The area growth rate of the dimming shows a steep increase within the first hour after the onset of the flare, and its peak (vertical solid green line) coincides in time with the peak of the time derivative of the flare SXR flux $\dot{F}$.

The area growth of the coronal dimming is related to the expansion of the associated CME. As a consequence, the impulsive phase of the dimming is determined by the evolution of the dimming area, and more specifically by the area growth rate $\dot{A}(t)$. Following the criteria in \citet{Dissauer2018b}, we set the end of the dimming impulsive phase to occur when the area growth rate $\dot{A}(t)$ decreases below 15\% of its maximum $\dot{A}_\text{max}$; that is when
\begin{equation}
\dot{A}(t) \leq 0.15\cdot \dot{A}_\text{max}.
\end{equation}
The end of the impulsive phase for event no. 29 is shown as vertical dashed green line in Fig.~\ref{fig:evolution-example}. In this example, the end of the impulsive dimming phase occurs soon after the end of the associated flare. For events that are followed by another $\geq$M1.0 flare within the considered analysis period (2~hr), the end of the impulsive phase was set to the start time of the subsequent flare. We also manually adjusted the end of the impulsive phase for three other events (nos. 9, 44, 46 in Table~\ref{table:flare_list}) where the ending of the impulsive dimming phase was misidentified before the flare end time.

The region of the dimming area where the magnetic flux density $B$ in the SDO/HMI LOS magnetograms surpasses the noise level ($|B|>10$~G) is defined as the `magnetic area' $A_\phi(t)$ and its time derivative is the magnetic area growth rate $\dot{A}_\phi(t)$. The magnetic area $A_\phi(t)$ signifies how much flux is covered by the cumulative dimming mask at time $t$ by projecting the dimming pixels identified in the AIA 211~\AA~images until time $t$ onto the co-registered HMI LOS magnetograms. From the evolution of the magnetic area, the cumulative unsigned magnetic flux $\phi(t)$, the positive flux $\phi_+(t)$ and the negative flux $|\phi_-(t)|$ are extracted, as well as their associated magnetic flux rates ($\dot{\phi}$ and $\dot{\phi}_\pm(t)$).  We also calculate the mean unsigned magnetic flux density $B_{\text{us}}(t)$, where the magnetic information is decoupled from the area of the dimming region.
For event no. 29, Fig.~\ref{fig:evolution-example} (d) shows the evolution of $\phi(t)$, $\phi_-(t)$, and $\phi_+(t)$ in black, red, and blue, respectively; panel (e) shows their magnetic flux rates $\dot{\phi}_-$ (red) and $\dot{\phi}_+(t)$ (blue). 

The evolution of the drop in brightness $I_\text{drop}(t)$\footnote{$I_\text{drop}(t)$ is equivalent to $I_\text{cu, diff}(t)$ in \citet{Dissauer2018b}} of the dimming region is calculated for a constant area set by the cumulative dimming pixel mask at the end of the impulsive phase. The mask is shown in cyan contours in the bottom middle panel of Fig.~\ref{fig:overview-example}.  $I_\text{drop}(t)$ is calculated as the sum of the brightness in base-difference units of the pixels within the dimming mask. Thus, it represents how much the intensity decreases with respect to the pre-event image within the dimming region. The brightness drop rate $\dot{I}_\text{drop}$; that is, how quickly the brightness decreases, is given by the derivative of the brightness drop. The mean intensity drop $\bar{I}_\text{drop}$ is also calculated as the brightness drop divided by the total dimming area to extract the area-independent brightness decrease.  Panel (c) in Fig.~\ref{fig:evolution-example} shows the brightness drop evolution, $I_\text{drop}(t)$ in base-difference units (black line) for flare no. 29 and its derivative $\dot{I}_\text{drop}$ (purple line). The dimming brightness strongly decreases as the dimming area grows and it reaches its minimum after the end of the flare. 

The values for the characteristic parameters are calculated with respect to the impulsive phase of the dimming. For $A$, $A_\phi$, $\phi$, $\phi_\pm$ and $B_\text{us}$ the parameter value is given by its cumulative sum at the end of the impulsive phase. For $I_\text{drop}$ and $\bar{I}_\text{drop}$ the minimum value within the impulsive dimming phase is taken. These are first order dimming parameters. For second order dimming parameters, $\dot{A}$, $\dot{A}_\phi$, $\dot{\phi}$, $\dot{I}_\text{drop}$, the parameter value is the maximum (minimum in the case of $\dot{I}_\text{drop}$). For further information on the method and the characteristic parameters, we refer to \citet{Dissauer2018a}.
In Table~\ref{table:dimming_properties} the brightness parameters listed are derived from base-difference images, showing the decrease in intensity with respect to the pre-event intensity level in the dimming region.

\subsection{Dimming selection criteria} \label{sec:methods:selection}

The dimming detection method (Sect.~\ref{sec:methods:detection}) is not intended to be used as a dimming identification tool, but as an analysis tool; that is, it assumes there is a dimming and extracts its properties. Nevertheless, in this study we applied it indiscriminately to all M- and X-class flares without prior knowledge of the presence of a coronal dimming. This approach was chosen in order to also check whether flares that are not associated with a CME (confined flares) may show dimming signatures. However, the movement of coronal loops or the disappearance of flare ribbons from a preceding flare might cause the detection algorithm to consider the changes as potential coronal dimmings even though they arise from different physical processes. The case of AR~13664 is particularly complex due to its substantial size and brightness, its high dynamics and intense flaring activity. In order to avoid false detections of processes unrelated to a mass depletion, we applied a thresholding technique based on the characteristic parameter distributions. 

\begin{figure}[t] 
	\centering
    \resizebox{\hsize}{!}{\includegraphics{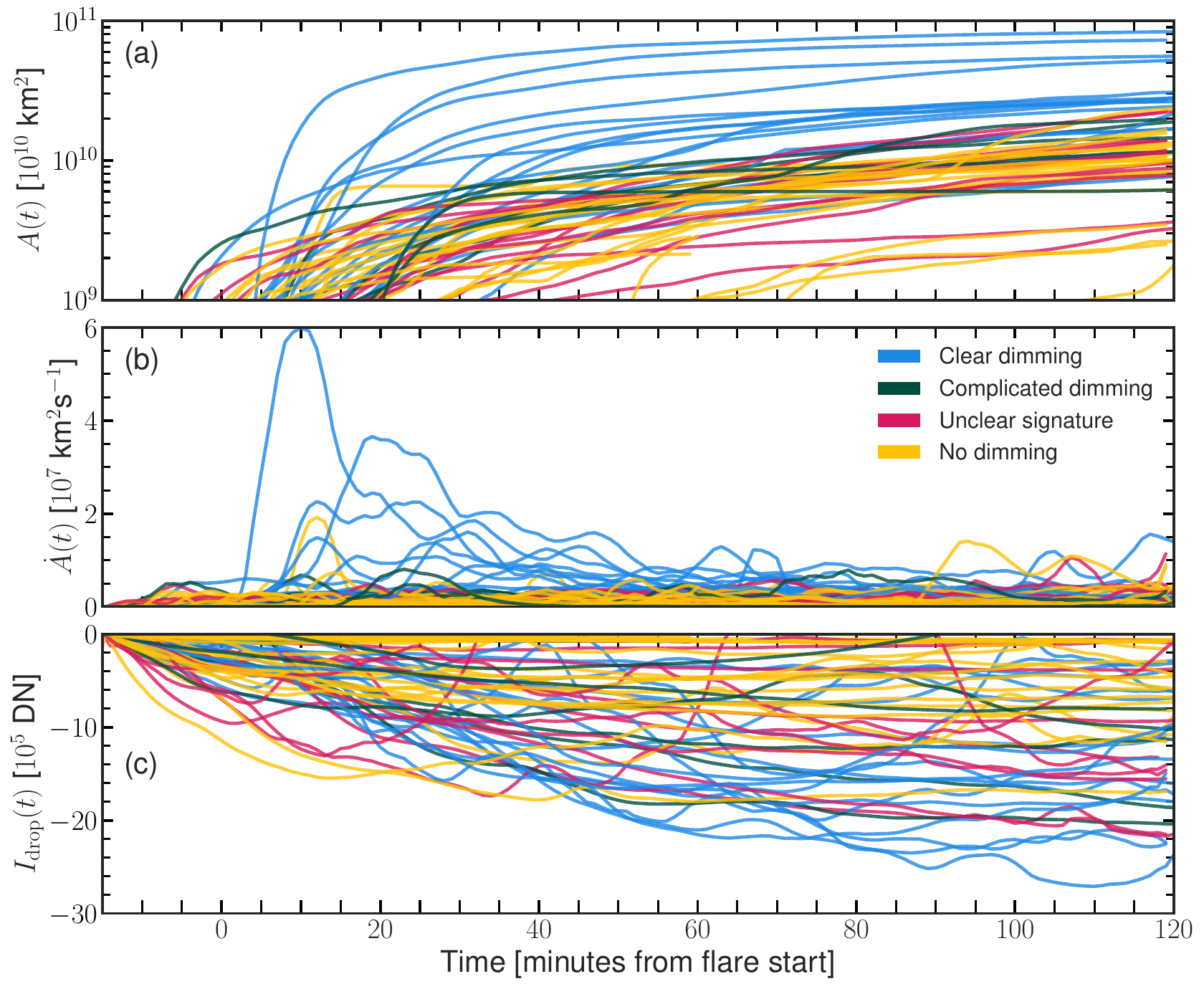}}
	\caption{Superposed epoch analysis of selected dimming parameters for all flares above M1.0 class from AR~13664 during 2024 May 1 to May 15. Events are colour-coded according to their visual characterisation as dimmings: blue for events with clear dimming signatures, green for dimming events that are detectable but difficult to interpret, pink for events with ambiguous brightness decreases, and yellow for events with no visible dimming. The panels show the time evolution of (a) dimming area $A(t)$, (b) dimming area growth rate $\dot{A}(t)$, and (c) brightness drop $I_\text{drop}(t)$.}
	\label{fig:selection}
\end{figure}

We first performed a superposed epoch analysis with the evolution of the main characteristic parameters for every event in order to investigate the differences between clear coronal dimmings and false detections. Fig.~\ref{fig:selection} shows the evolution of the dimming area $A(t)$ (a), area growth rate $\dot{A}(t)$ (b) and brightness drop $I_\text{drop}(t)$ (c) for all the events from AR~13664. We colour-coded them based on their visual dimming identification. If a dimming is visually easily identifiable (such as no. 29 in Fig.~\ref{fig:overview-example}) it is shown in blue; if a dimming can be observed but it has a more complex structure and difficult characterisation, it is plotted in green; unclear signatures where a faint darkening is observed but the identification is doubtful are shown in magenta; and detections in which a dimming is unrecognizable are plotted in yellow. Clear dimmings (blue lines) have a very distinct shape in $A(t)$ and $\dot{A}(t)$: they show a fast impulsive growth start as indicated by the early peak in the area growth rate (panel (b)), and a larger area throughout the first 2~hr after the start of the associated flare (panel (a)). In contrast, distinguishing clear coronal dimmings from false detections is challenging when looking at the brightness decrease $I_\text{drop}$ in panel (c). This is also the case for the rest of characteristic parameters as illustrated in Appendix~\ref{Appendix_epoch}.

Therefore, to avoid analysing unrelated darkenings we only considered detections for which the average area growth rate within the first hour after the flare start is above $1.8\times10^6$~km$^2$~s$^{-1}$; that is:
\begin{equation}
    \overline{\left(\frac{dA}{dt}\right)}_{1\text{h}} > 1.8 \times 10^6 \, \text{km}^2\text{s}^{-1}.
\end{equation}
This is equivalent to selecting dimmings that grow by $\Delta A \geq 6.48\times10^9$~km$^2$ within the first hour after the flare start time.

We cross-checked this threshold with the parameter distributions found in \citet{Dissauer2018b}. They found that, among dimmings associated with B, C, M, and X-class flares, dimming areas reached on average values of $2.7\times10^{10}$~km$^2$ and the dimming duration was around 1~hour. Thus, most of the area growth occurs within the first hour and the average area growth rate within that time is $7.5\times10^6$~km$^2$~s$^{-1}$, which is higher than the minimum threshold set in this study.

After applying this dimming selection criteria, we manually removed one false dimming. The coronal dimming detection for the M2.7 flare on May 2 at 20:52~UT does not correspond to a dimming; that is, to a decrease in coronal brightness due to a depletion of plasma, but to a darkening of the image caused by cool ejecta.

\subsection{Statistical analysis} 
\label{sec:methods:statistical_analysis}
The uncertainties in the characteristic parameters were calculated following the method used in \citet{Dissauer2018a} and \citet{chikunova2020coronal}. We performed the dimming detection with a threshold 5\% higher and 5\% lower than the optimal threshold of $-0.19$, and calculated the respective dimming parameters. From these detections we extracted the mean and the standard deviation, as reflected in Fig.~\ref{fig:evolution-example} where the shaded region represents the $1\sigma$ standard deviation. 

To analyse the statistical relationship between the characteristic dimming parameters, the basic flare quantities and the CME maximum speed, we calculated the Pearson correlation coefficient ($c$) in log-log space. The errors in $c$ were calculated with a bootstrapping method, where N-out-of-N random data pairs with replacement are selected to calculate $c$ over 10,000 times, and the mean and standard deviation $\bar{c}\pm\Delta c$ are extracted \citep{Wall_Jenkins_2012}. The qualitative strength of the correlation is described following \citet{Kazachenko2017database}: $c=[0.2, 0.4] -$weak; $c=[0.4, 0.6] -$moderate; $c=[0.6, 0.8] -$strong; $c=[0.8, 1.0] -$very strong. 

This statistical analysis was compared to that performed by \citet{Dissauer2018b, Dissauer2019}, where the analysis and detection methods are identical. Their study focused on 62 eruptive events observed when the STEREO spacecraft were positioned in quasi quadrature with the Sun-Earth line between 2010 and 2012. Thereby, they studied coronal dimmings on-disc with eruption sites within $40^\circ$ from the solar disc centre using SDO data while simultaneously analysing CMEs observed close to the limb by STEREO-A and/or -B, minimising projection effects. The studied flares had GOES classes ranging from B to X.
This approach allowed us to set the characteristics of coronal dimmings from a single very flare-productive active region, AR~13664, into context with a general distribution of dimming parameters from a variety of different ARs and associated with flares of different classes.

\section{Results} \label{sec:results}
Table~\ref{table:flare_list} summarises all the M and X-class flares that originated from AR~13664, along with the associated CMEs and coronal dimmings. In total, these are 67 major flares (12 X-class, 55 M-class). The CME speed listed is the maximum speed derived from the velocity-height curves in the CDAW SOHO/LASCO catalogue. Sixteen flares are accompanied by an on-disc coronal dimming, six by an off-limb dimming (last column in Table~\ref{table:flare_list}), while 45 flares show no dimming at all. The characteristic parameters of all the on-disc dimmings derived as described in Sect.~\ref{sec:methods} is summarised in Table~\ref{table:dimming_properties}.

\subsection{Flare-CME-dimming associations}
During the period from May 1 to May 15, AR~13664 produced 23 CMEs: ten of which were halo CMEs, five were partial halo CMEs, and eight exhibited an angular width of less than $140^\circ$.
Among these CMEs, ten are associated with X-class flares, while the remaining 13 are linked to M-class flares. This corresponds to 83\% of X-class flares being associated with CMEs, compared to only 23\% of M-class flares. This is substantially smaller than for a general flare population, where it is found that about 60\% of the M-class flares are associated with a CME \citep{Yashiro2006different}. This difference hints at an above average magnetic confinement of the large AR~13664. High magnetic flux and free energy increase the likelihood of confined flares \citep{Li2021magnetic}. With magnetic flux values up to $1.9\times10^{23}$~Mx  \citep{Jarolim2024magnetic,hayakawa2024solar}, AR~13664's low eruptivity may be explained by its strong magnetic confinement.

Of the 23 CMEs associated with AR~13664, six are accompanied by off-limb dimmings, all observed above the western limb. Except for two events, these off-limb dimmings follow X-class flares. Additionally, 13 CMEs are associated with on-disc dimmings, six of which occur with X-class flares. Notably, the two X-class flares not linked to a CME do not exhibit a corresponding dimming event either.

Two of the CMEs that lack an associated dimming occurred on May 2 and May 3, when AR~13664 was positioned near the Eastern limb (beyond $60^\circ$). At this time, the AR was in an early growth phase, and these CMEs were associated with M2.7-class flares. As a result, the expected dimming signatures would be relatively weak and difficult to detect due to projection effects.

Another CME unaccompanied by coronal dimming is associated with an M5.9-class flare occurring at 13:58 UT on May 10. However, at the time of this event, significant magnetic restructuring is observed in the southern hemisphere, outside of AR~13664. This large-scale reconfiguration may have been the dominant factor driving the CME, which would indicate that the CME is not directly associated with the AR, thus explaining the absence of a detectable dimming signature.

Furthermore, \citet{hayakawa2024solar} do not include these three CMEs in their analysis of `major' CMEs from AR~13664. Event no. 62 is the only CME in the event list by \citet{hayakawa2024solar} that does not have an associated dimming. The flare itself is not included in the GOES flare catalogue because the peak occurs within the rising phase of the X1.7 flare right after (event no. 63). 
Additionally, 3 dimming events are not linked to any CME: an M1.0-class flare on May 5 at 18:34 UT, an M2.0-class flare on May 8 at 19:15 UT, and an M2.3-class flare on May 9 at 06:03 UT.

We also note that events no. 14 and 15 in Table~\ref{table:flare_list} are associated with the same CME, but only the latter one has a dimming. Thus, we associate the characteristic dimming parameters of event no. 15 with the CME, even though the onset of the CME is most probably cotemporal to the flare event no. 14.

Among the on-disc dimming events, six are associated with X-class flares, while ten are linked to M-class flares. Similarly, among the off-limb dimming events, four are associated with X-class flares, whereas two are connected to M-class flares. Consequently, 83\% of X-class flares exhibit an associated coronal dimming, while 22\% of M-class flares are accompanied by a dimming.

\subsection{Relations between flare and dimming parameters} \label{sec:results:flare}
\begin{figure}
    \begin{subfigure}{\columnwidth}
        \centering
        \resizebox{\hsize}{!}{\includegraphics{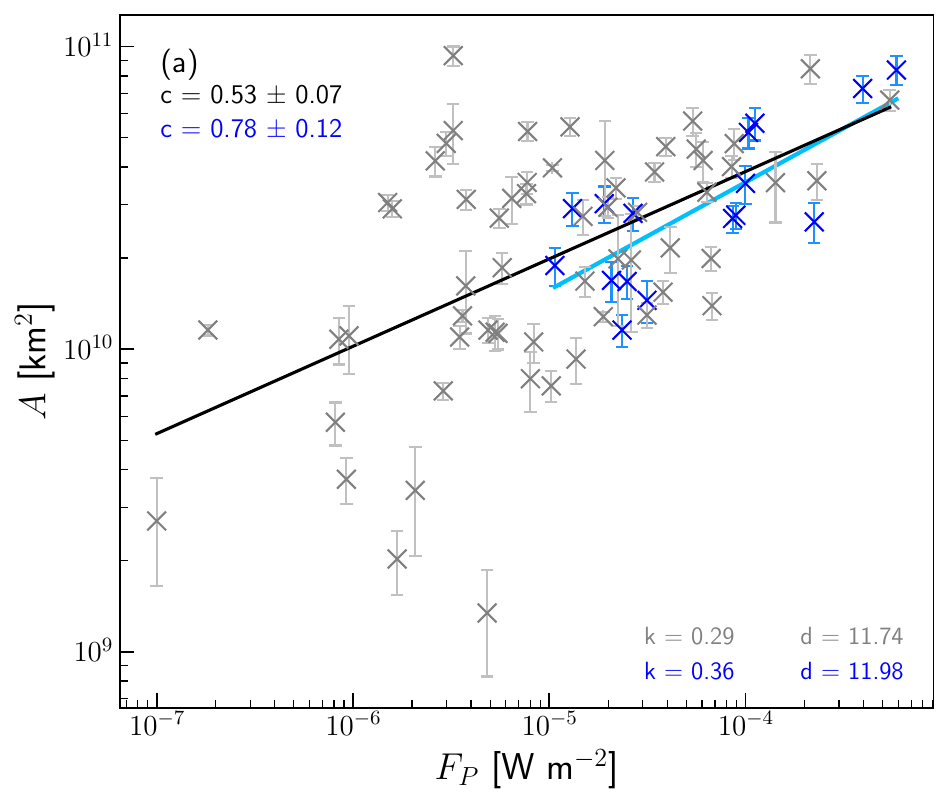}}
    \end{subfigure}
    \hfill
    \begin{subfigure}{\columnwidth}
        \centering
        \resizebox{\hsize}{!}{\includegraphics{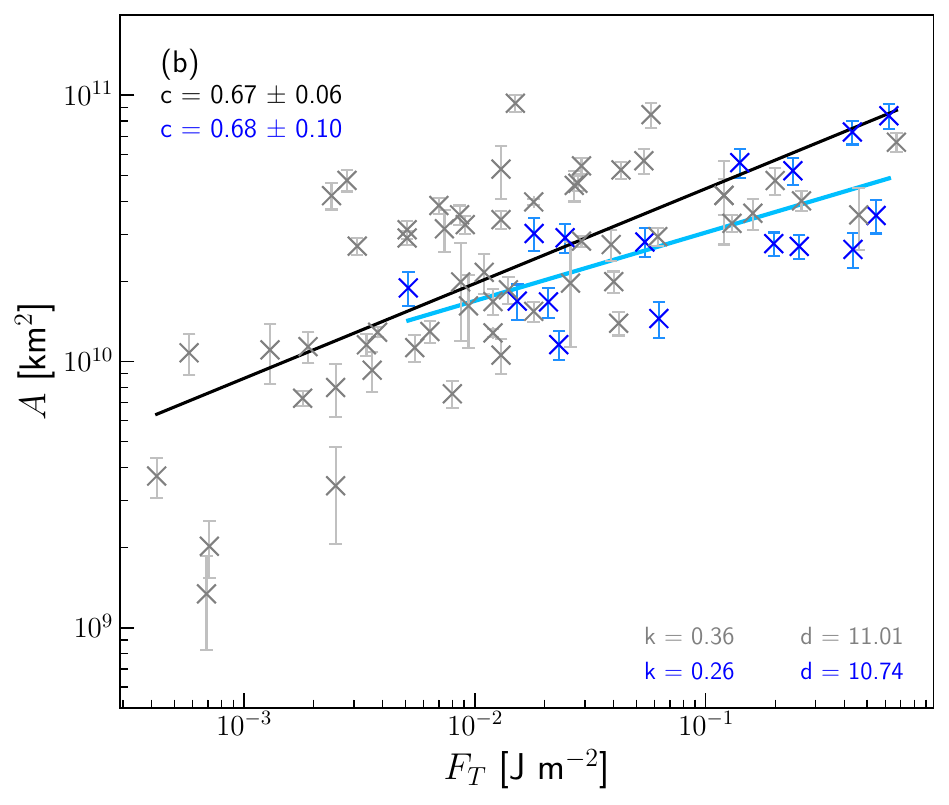}}
    \end{subfigure}
    \caption{Dimming area $A$ against (a) the GOES SXR peak flux $F_P$ and (b) the GOES SXR fluence $F_T$. Blue crosses represent dimmings from the May 2024 events, while grey crosses correspond to dimming events from \citet{Dissauer2018b}. The black (blue) regression lines are fitted exclusively to the grey (blue) data points.}
    \label{fig:cor:A_Fp_Ft}
\end{figure}

\begin{figure}[t] 
	\centering
    \resizebox{\hsize}{!}{\includegraphics{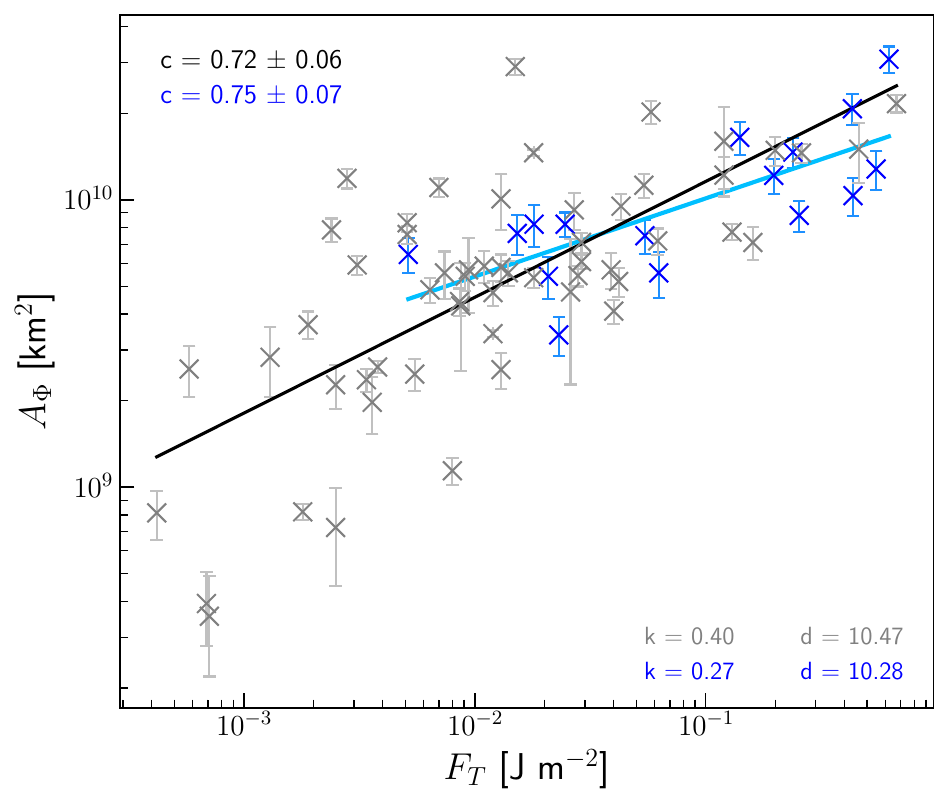}}
	\caption{Same as Fig.~\ref{fig:cor:A_Fp_Ft} but for the magnetic dimming area $A_\phi$ against the GOES SXR fluence $F_T$.}
	\label{fig:cor:amag_ft}
\end{figure}

\begin{figure}[t] 
	\centering
    \resizebox{\hsize}{!}{\includegraphics{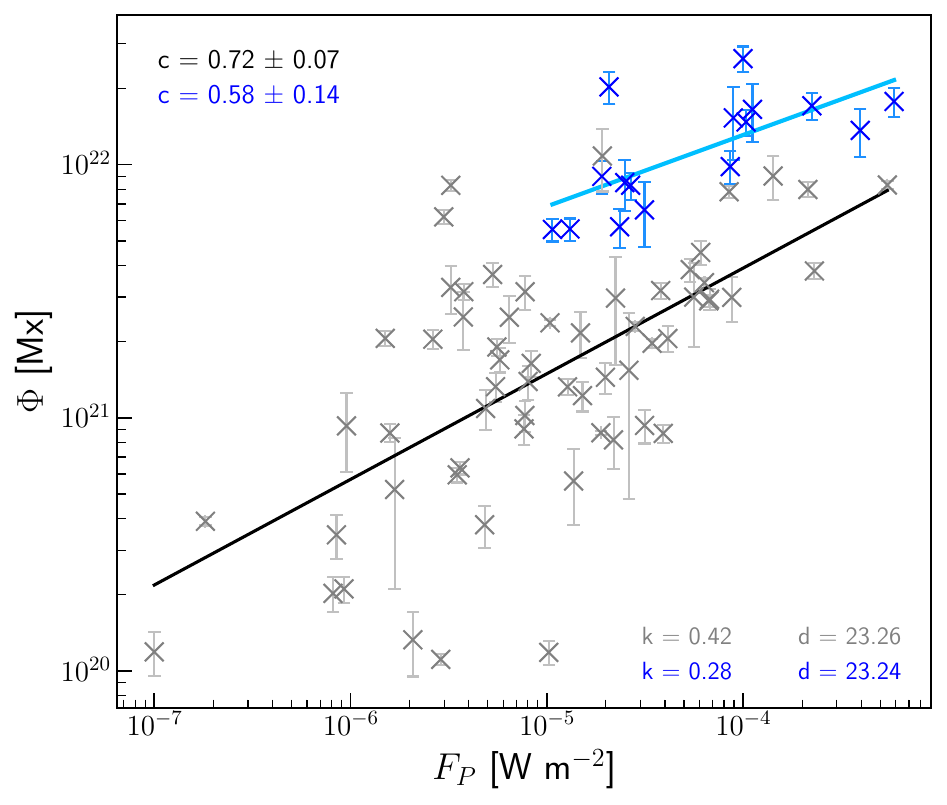}}
	\caption{Same as Fig.~\ref{fig:cor:A_Fp_Ft} but for the total unsigned magnetic flux $\phi$ against the GOES SXR peak flux $F_P$.}
	\label{fig:cor:flux_goes}
\end{figure}

Figures \ref{fig:cor:A_Fp_Ft}--\ref{fig:cor:vMas_A_dA_true} show the most significant dependencies among characteristic dimming parameters, flare parameters and CME maximum velocity. In each plot, grey crosses represent data adapted from \citet{Dissauer2018b}, while blue crosses correspond to the May 2024 events analysed in this study, unless differently stated. In what follows, results and data from \citet{Dissauer2018b} will be denoted as KD18. The black (blue) line in the figures depicts the linear regression fit to the distribution of KD18 (May 2024 events) data in log-log space with 
\begin{equation}
    \log{Y}=k\cdot\log{X}+d,
\end{equation}
where $k$ and $d$ are the regression line coefficients. The coefficients are indicated in the figures in grey and blue for KD18 and the May 2024 events data, respectively. The correlation coefficients $c=\bar{c}\pm\Delta c$ for KD18 (May 2024 events) data are indicated in black (blue) in the upper left corner of the figures. The correlation coefficients for all the parameters are detailed in Fig.~\ref{fig:app:correlation} of Appendix~\ref{Appendix_correlations}. 

To understand the relationship between the flares and their associated coronal dimmings, we correlate the characteristic dimming parameters with the GOES SXR peak flux $F_P$, the peak of the flux derivative $\dot{F}_P$, and the flare fluence $F_T$. The flare parameters for each event are shown in Table~\ref{table:flare_list}.  Figures \ref{fig:cor:A_Fp_Ft}--\ref{fig:cor:flux_goes} show a selection of correlation plots between flare and dimming parameters compared to KD18 data. In Fig.~\ref{fig:cor:A_Fp_Ft} we plot the total dimming area $A$ as a function of peak flare flux $F_P$ (panel (a)) and flare fluence $F_T$ (panel (b)). The dimming areas for the May 2024 events fall within the observed distribution of data from KD18, flares with higher peak and fluence being accompanied by larger dimmings. The correlation between $A$ and $F_P$ is larger for the May 2024 events dimmings ($c=0.78\pm 0.12$) than for KD18 data ($c=0.53\pm0.07$).  This is also true for the magnetic area $A_\phi$, where the largest -- very strong -- correlation between flare parameters and dimming parameters ($c=0.84\pm0.08$) is found.
The dimming area also shows strong correlation with flare fluence ($c=0.68\pm0.10$), while a weak correlation is found with the flare flux derivative peak $\dot{F}_P$ ($c=0.38\pm0.19$).  

The $p$-values of the correlation coefficients are in general lower for KD18 events compared to May 2024 events due to the larger sample size: 62 events in KD18 compared to 16 dimming events in May 2024. For example, the correlation coefficient between $A$ and $F_P$ has a $p$-value of $9\times10^{-6}$ for KD18 events and of $4\times10^{-4}$ for May 2024 events. The smaller sample size is also reflected in the uncertainty ranges for the correlation coefficient $c$ calculated with the bootstrapping method described in Sect.~\ref{sec:methods:statistical_analysis}, with larger values indicating less significance. We refer to Fig.~\ref{fig:app:correlation} for an overview of the statistical significance of each correlation coefficient we calculated. In what follows, all significant correlation coefficients mentioned for the May 2024 events dimmings have a $p$-value below 0.02. Thus, while these coefficients might be more statistically significant for the KD18 sample, they are also significant for the May 2024 events.

Figure~\ref{fig:cor:amag_ft} shows the dimming magnetic area $A_\phi$ with respect to the flare fluence $F_T$. The dimmings from the May 2024 events exhibit a similar trend to those observed in KD18, with a strong correlation between the two parameters ($c=0.75\pm0.07$). This positive correlation holds true for the total unsigned magnetic flux  ($c=0.68\pm0.18$) and the dimming brightness drop ($c=0.45\pm0.16$). These findings align with the conclusions of \citet{Dissauer2018b}, which suggested that higher flare fluence corresponds to a bigger magnetic dimming area, increased magnetic flux, and a darker dimming region. Notably, the correlation between $A_\phi$ and $F_T$ is larger in the May 2024 events, likely due to the magnetic structures involved being similar in the May 2024 events and thus strengthening the connection between the magnetic area and the amount of energy released. 

Figure~\ref{fig:cor:flux_goes} shows the total unsigned magnetic flux $\phi$ with respect to the peak flare flux $F_P$. The May event values for $\phi$ fall above the regression line for KD18, and show a correlation of $c=0.58\pm0.14$. Thus, the May 2024 events show in general higher total unsigned magnetic fluxes $\phi$ for the same flare class range considered. Since in the relation to the dimming areas no such difference appears, it is clear that the difference results from the stronger underlying magnetic flux density. In addition, since the dimmings in this study are detected considering only M and X-class flares, the relation of $\phi$ and $F_P$ can only be considered for a subregion ($10^{-5}<F_P<10^{-3}$) of the total flare class range which may cause the correlation to be smaller. 

The peak flare flux $F_P$ also shows a strong correlation with the brightness drop rate $\dot{I}_\text{drop}$ ($c=0.70\pm0.09$), and a moderate correlation with the brightness drop $I_\text{drop}$ ($c=0.41\pm0.14$). However, this correlation becomes negligible when we consider the area-independent parameters $B_\text{us}$ and $\bar{I}_\text{drop}$ (see Appendix~\ref{Appendix_correlations}), indicating that the main correlation of the flare peak flux with the dimming parameters results primarily from the dimming area $A$. 

Overall, the May 2024 events dimmings show an enhanced correlation with the peak flare flux $F_P$ and flare fluence $F_P$, compared to the analysis by \cite{Dissauer2018b}. This can be attributed to the dimmings originating from a single AR and the flare class analysed being higher on average. In addition, the total unsigned magnetic flux $\phi$ exceeds the values of the general dimming population in KD18 because of the specially strong underlying magnetic fields in AR~13664.

\subsection{Correlations of characteristic dimming parameters} \label{sec:results:dimming}

\begin{figure}[t] 
	\centering
    \resizebox{\hsize}{!}{\includegraphics{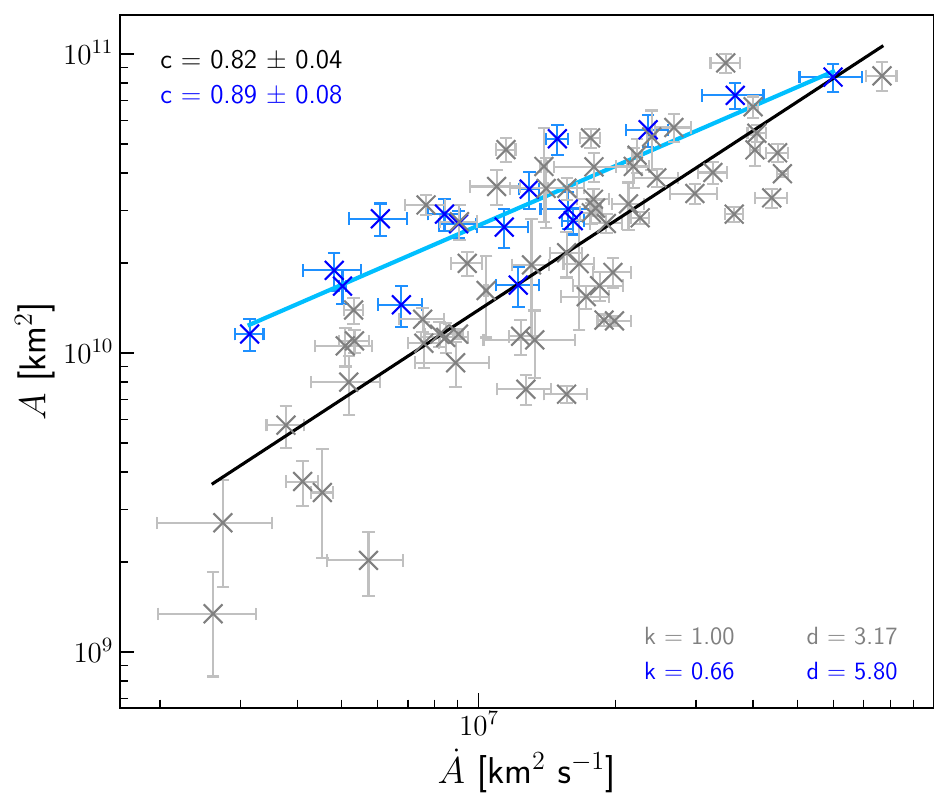}}
	\caption{Same as Fig.~\ref{fig:cor:A_Fp_Ft} but for the dimming area $A$ and the area growth rate $\dot{A}$.}
	\label{fig:cor:area_growth}
\end{figure}

\begin{figure}
     \begin{subfigure}{\columnwidth}
        \centering
        \resizebox{\hsize}{!}{\includegraphics{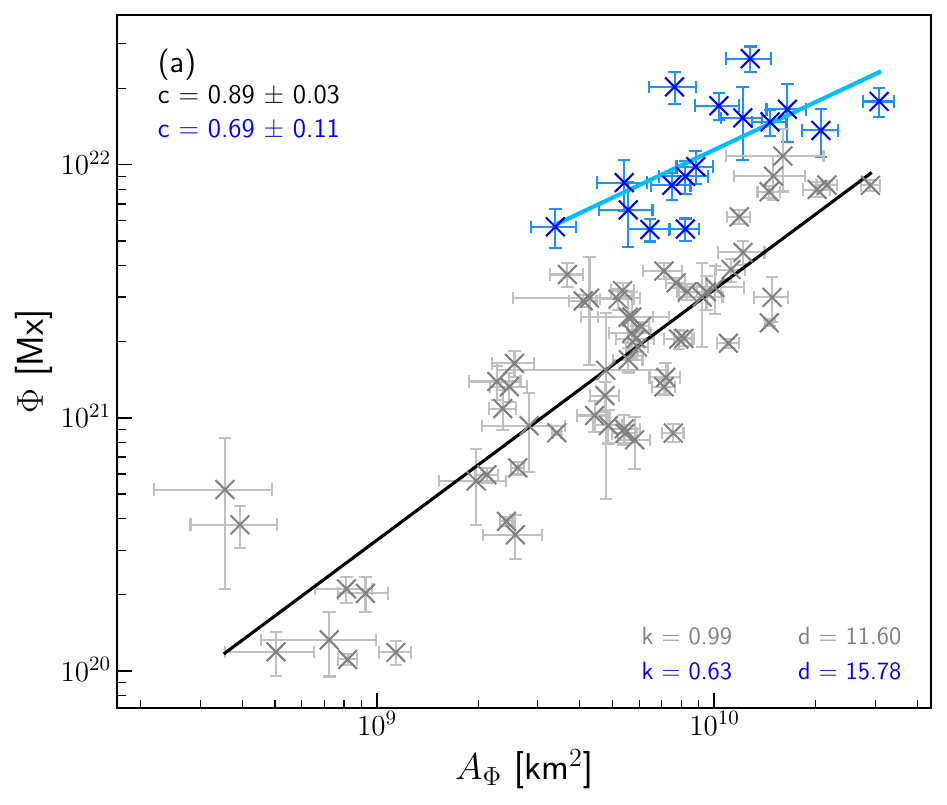}}
    \end{subfigure}
    \hfill
    \begin{subfigure}{\columnwidth}
        \centering
        \resizebox{\hsize}{!}{\includegraphics{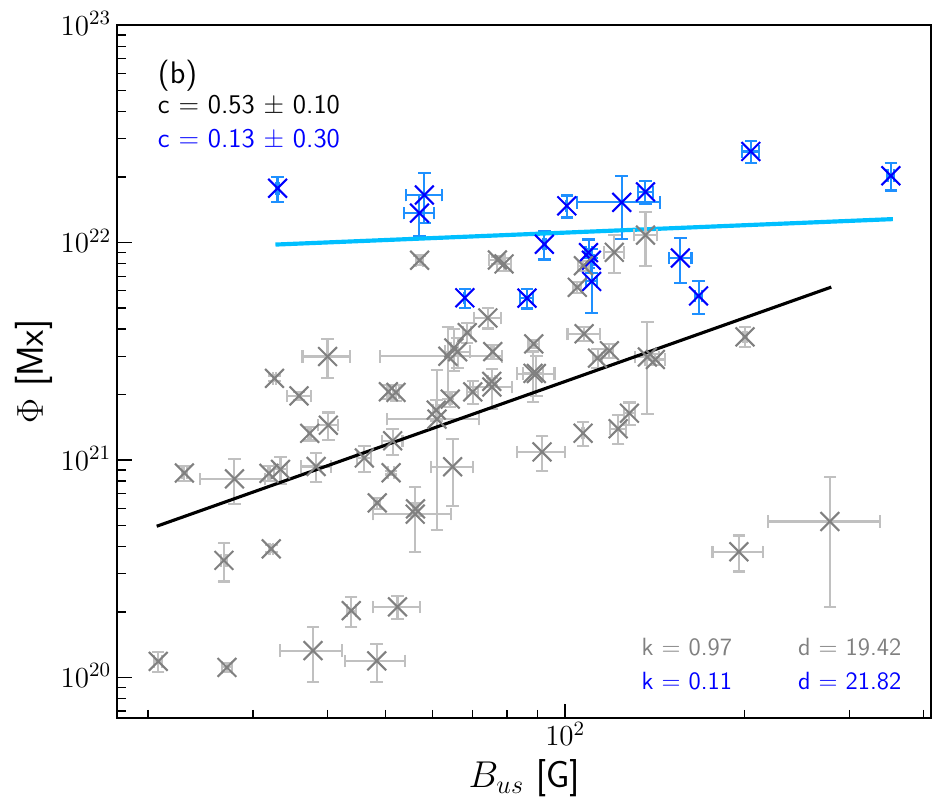}}
    \end{subfigure} 
    \caption{Same as Fig.~\ref{fig:cor:A_Fp_Ft} but for the total unsigned magnetic flux $\phi$ against (a) the magnetic dimming area $A_\phi$ and (b) the mean unsigned magnetic flux density $B_\text{us}$.}
    \label{fig:cor:flux_amag_B}
\end{figure}

\begin{figure}[t] 
	\centering
    \resizebox{\hsize}{!}{\includegraphics{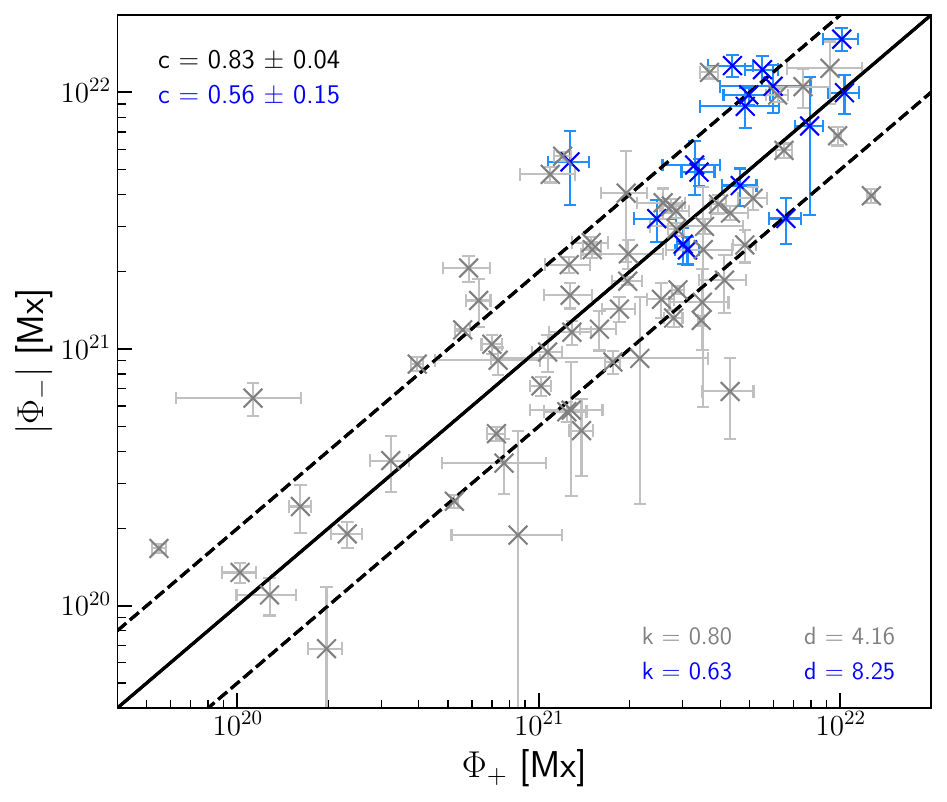}}
	\caption{Same as Fig.~\ref{fig:cor:A_Fp_Ft} but for the negative magnetic flux $|\phi_-|$ and the positive magnetic flux $\phi_+$. The black line indicates the 1:1 correspondence, and the dashed lines indicate ratios of 0.5 and 2.}
	\label{fig:cor:fluxneg_fluxpos}
\end{figure}

\begin{figure}[t] 
	\centering
    \resizebox{\hsize}{!}{\includegraphics{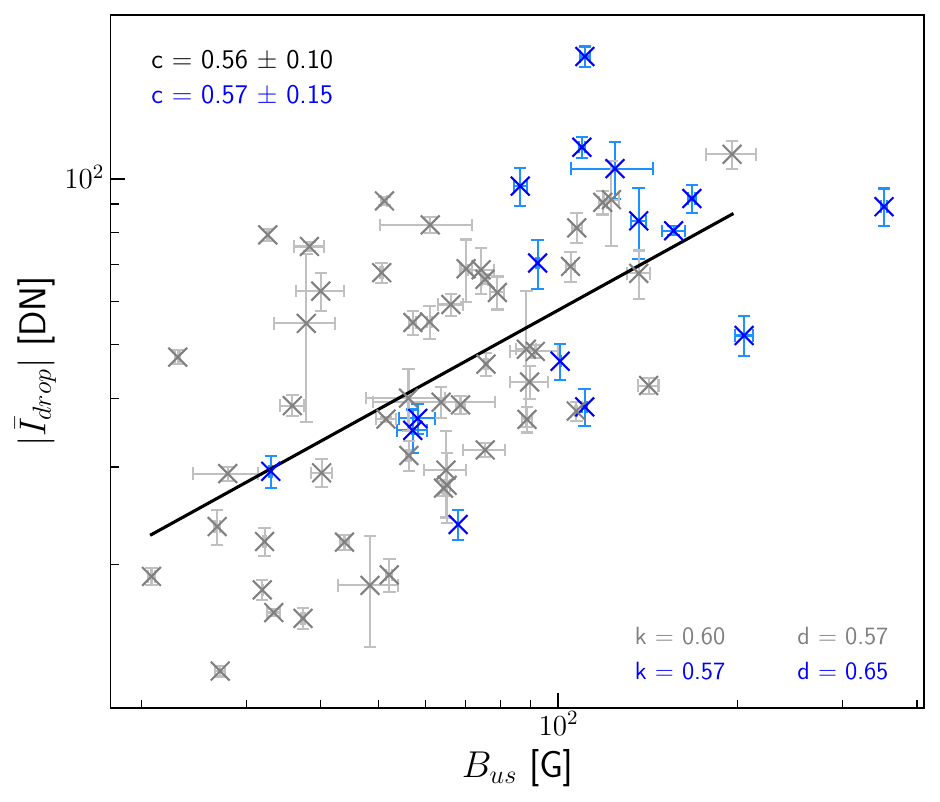}}
	\caption{Same as Fig.~\ref{fig:cor:A_Fp_Ft} but for the mean brightness drop $|\bar{I}_\text{drop}|$ and the mean magnetic flux density $B_\text{us}$.}
	\label{fig:cor:Imean_B_A}
\end{figure}

Figures \ref{fig:cor:area_growth}--\ref{fig:cor:Imean_B_A} present the most significant correlations among the dimming parameters. Figure~\ref{fig:cor:area_growth} shows the relationship between the total area $A$ and the maximum area growth rate $\dot{A}$. The dimming area values cover a range from $1.16\times10^{10}$~km$^2$ to $8.38\times10^{10}$~km$^2$, which is within the values found by \citet{aschwanden2017global} and \citet{Dissauer2018b}. The May 2024 events dimmings consistently exhibit a moderately larger total area per maximum area growth rate compared to KD18, which implies that for events with the same impulsivity, the dimmings observed in AR~13664 were larger than expected. This can be attributed, on one hand, to the AR being larger than average, and on the other hand, to the fact that KD18 also contains weaker flares whereas for the May 2024 events we focus only on M- and X-class flares. Given the correlation between flare class and dimming area (see Sect.~\ref{sec:results:flare}) for the May 2024 events we exclusively sample dimmings with larger areas. 

Figure~\ref{fig:cor:flux_amag_B} (a) illustrates the total unsigned magnetic flux $\phi$ in relation to the magnetic area $A_\phi$. All of the May 2024 events dimmings have a total unsigned magnetic flux above $5.5\times 10^{21}$~Mx, which is in contrast to KD18 data where 90\% of the dimmings had $\phi< 5.0\times10^{21}$~Mx. The magnetic areas are also situated at the upper range of the values in KD18 because the dimming areas are larger. For a specific $A_\phi$, the unsigned magnetic flux is significantly larger compared to the fit derived from KD18. Similarly to the dimming area, this is due partly because we are considering only the higher energy flares, and partly because the magnetic flux within AR~13664 is considerably higher than that of the majority of ARs \citep{hayakawa2024solar}. As mentioned in Sect.~\ref{sec:results:flare}, the dimming magnetic flux is larger in relation to flare class compared to what is presented in KD18 (see Fig.~\ref{fig:cor:flux_goes}); and the mean magnetic flux density in the dimming region for the May events is, except for three cases, above 65~G, while in KD18 data fewer than half of the events have flux densities above 65~G. Therefore, we conclude that the systematically large magnetic dimming flux is mainly given by the strong underlying magnetic flux density.

The large magnetic flux density can also be seen in Fig.~\ref{fig:cor:flux_amag_B} (b) where the total unsigned magnetic flux $\phi$ is presented against the magnetic flux density $B_\text{us}$. The data from KD18 indicated a smaller correlation between $\phi$ and $B_\text{us}$ compared to the correlation between $\phi$ and $A_\phi$, leading \citet{Dissauer2018b} to conclude that the differences in $\phi$ stem mainly from changes in magnetic dimming area rather than from the strength of the underlying magnetic field. This conclusion is more pronounced in the May 2024 events, as $\phi$ and $B_\text{us}$ show no correlation while $\phi$ and $A_
\phi$ have a correlation of $c=0.69\pm0.11$. Furthermore, an anticorrelation of $c=-0.60\pm0.20$ is seen between $B_\text{us}$ and $A_\phi$ (see Appendix~\ref{Appendix_correlations}), which can be attributed to the expansion of the magnetic dimming area beyond the dense core of the AR into regions with weaker magnetic fields, leading to a decrease in $B_\text{us}$ as it is averaged over the area. This anticorrelation is specifically observed in this study because we are looking at dimmings within the same AR and it does not appear in KD18 data (see Fig.~\ref{fig:app:anticorrelation}), as the active regions for the general dimming population are more diverse.

Figure~\ref{fig:cor:fluxneg_fluxpos} shows the negative magnetic flux against the positive magnetic flux covered by the dimming regions. The 1:1 correspondence, indicating that the flux of the dimming regions is perfectly balanced, is given as a black line. \citet{Qiu2005magnetic} defined the range for balanced reconnection fluxes for flare observations to be between 0.5 and 2.0, because measurement uncertainties hinder perfect balance detections. The dashed black lines in Fig.~\ref{fig:cor:fluxneg_fluxpos} indicate this region. The May 2024 events dimmings exhibit balanced fluxes analogous to those in KD18, but with the flux values lying at the higher end of the KD18 range.

Figure~\ref{fig:cor:Imean_B_A} (a) compares the mean brightness drop in the dimming area $|\bar{I}_{\text{drop}}|$ with the mean magnetic flux density $B_\text{us}$, both area-independent parameters. The mean brightness drop is plotted in absolute units, thus illustrating how deep the brightness decrease is. The two parameters show a moderate correlation ($c=0.57\pm0.15$), similar to that found in KD18. This correlation indicates that the stronger the underlying field strength, the greater the brightness drop. The brightness drop $I_\text{drop}$ has values ranging from $-27.10\times10^5$ to $-5.07\times10^5$~DN, which is within the typical values found in \citet{Dissauer2018b} and corresponds to a intensity decrease of 43.6\% to 73.0\% with respect to a pre-event image. The brightness drop rate $\dot{I}_\text{drop}$ for the May 2024 events is also within the expected values, ranging from $-0.39\times10^3$~DN~s$^{-1}$ to $-2.76\times10^3$~DN~s$^{-1}$. 

Altogether, the May 2024 events are associated with larger dimming areas than expected from their area growth rate. Positive and negative magnetic fluxes are balanced, consistent with previous findings. The total unsigned magnetic flux is larger compared to the magnetic dimming area than in \cite{Dissauer2018b}, which is a reflection of the strong underlying magnetic fields. The total magnetic flux correlates with dimming area but not with magnetic flux density, which indicates a stronger dependency with the dimming area than the underlying field density. A moderate correlation between brightness drop and magnetic field strength indicates that darker dimmings occur in stronger field regions.

\subsection{Relations between dimming parameters and CME velocity} \label{sec:results:cme}
\begin{figure*}[t] 
    \centering
    \resizebox{\hsize}{!}{\includegraphics{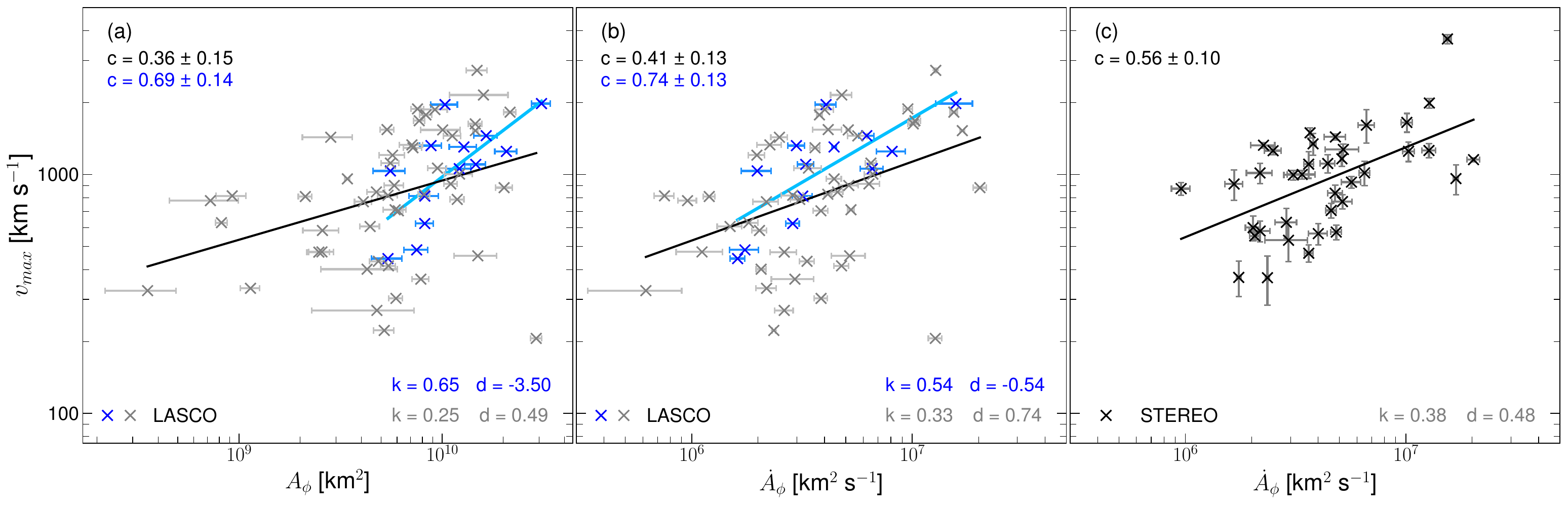}}
        \caption{Correlation plots of the CME maximum velocity $v_\text{max}$ derived from CDAW SOHO/LASCO C2 and C3 measurements against (a) the magnetic dimming area $A_\phi$ and (b) the magnetic dimming area growth rate $\dot{A}_\phi$. Blue crosses represent dimmings from the May 2024 events, while grey crosses correspond to dimming events from \citet{Dissauer2019}. Panel (c) shows the maximum CME velocities for events adapted from \citet{Dissauer2019} derived from STEREO EUVI+COR1+COR2 measurements. The black (blue) regression lines are fitted exclusively to the grey (blue) data points.}
    \label{fig:cor:vMas_A_dA_true}
\end{figure*}

Among the 16 on-disc dimming events, 13 are associated with CMEs. To investigate the relationship between CMEs originating from AR~13664 and their associated on-disc dimming characteristics, we compare the maximum CME velocity $v_\text{max}$ determined from the LASCO coronagraphs with dimming characteristic parameters in the context of the results from \citet{Dissauer2019}.  Fig.~\ref{fig:cor:vMas_A_dA_true} shows $v_\text{max}$ with respect to the magnetic dimming area $A_\phi$ (panel (a)) and the magnetic dimming area growth rate $\dot{A}_\phi$ (panel (b)). The analysis reveals a positive correlation with both parameters, with correlation coefficients of $c=0.69\pm0.14$ for $A_\phi$ and $c=0.74\pm0.13$ for $\dot{A}_\phi$. 

In the study by \citet{Dissauer2019}, CME properties were determined using multi-instrument observations from low in the corona using STEREO/EUVI out to the coronagraphic field-of-view observed by COR1, and COR2 to identify the leading edge when the STEREO spacecraft were in quasi quadrature with SDO. This instrument combination allowed them to reconstruct the impulsive CME acceleration profile that typically occurs below 1--2 solar radii, i.e. below the coronagrapohic field-of-view. Further, in this study the dimmings were observed on-disc by SDO while the CMEs were observed near the limb by STEREO, which served to reduce projection effects in the derived CME kinematics.
Consequently, the reported values of $v_\text{max}$ in that study reflects the initial CME acceleration phase during its early evolution. To enable a direct comparison with the May 2024 events analysed here, we extract CME maximum velocities for the events in \citet{Dissauer2019} from the SOHO/LASCO CDAW catalogue in the same manner as for the May 2024 events.

When applying the correlation calculation to these CME velocities, we obtain correlation coefficients of $c=0.36\pm0.15$ for $A_\phi$ and $c=0.41\pm0.13$ for $\dot{A}_\phi$. The grey crosses in Fig.~\ref{fig:cor:vMas_A_dA_true} (a, b) represent the LASCO-based maximum velocities for the events studied in \citet{Dissauer2019}, showing that the May 2024 events exhibit a stronger correlation between LASCO $v_\text{max}$ and both magnetic dimming area and magnetic area growth rate.

However, the correlations derived from LASCO measurements for the events in \citet{Dissauer2019} are weaker than those obtained using the STEREO multi-instrument quadrature observations. For instance, \citet{Dissauer2019} report a correlation of $c=0.68\pm0.08$ between mean brightness drop $|\bar{I}_\text{drop}|$ and $v_\text{max}$ using velocities derived from STEREO observations, but this correlation decreases to $c=0.25\pm0.15$ when using LASCO observations. A similar trend is observed for the correlation with the magnetic flux rate $\dot{\phi}$, which decreases from  $c=0.6\pm0.1$ to $c=0.31\pm0.13$ when using LASCO data instead of STEREO data (see Figs.~\ref{fig:app:correlation_vmax_dphi} and \ref{fig:app:correlation_vmax_Imean} in Appendix~\ref{Appendix_correlations}).  This discrepancy arises because STEREO observations (with EUVI) track CMEs from their initiation at the solar surface up to 15~$R_\odot$, while LASCO observations begin at 2~$R_\odot$ and extend to 32~$R_\odot$, thereby missing the initial acceleration of the CME. 

Figure~\ref{fig:cor:vMas_A_dA_true} (c) illustrates this effect by showing $v_\text{max}$ as given in \citet{Dissauer2019} (derived from STEREO measurements) plotted against $\dot{A}_\phi$. Compared to panel (b), where all data originate from LASCO, STEREO-based measurements exhibit less scatter and a more pronounced correlation of $c=0.56\pm0.10$ rather than $c=0.41\pm0.13$. 

The reconstruction of the velocity profiles is influenced by the viewpoint of LASCO. Specially for Earth-directed CMEs, LASCO measurements -- taken from the Sun-Earth line -- are susceptible to significant projection effects that reduce the accuracy of derived velocities which could be reflected in the correlation calculations using only LASCO-based data. This limitation applies to the May 2024 events. Although STEREO-A provides additional observations for the May 2024 events, its position only $12^\circ$ ahead of Earth offered limited improvement. However, \citet{Dissauer2019} reported that 93\% of the CMEs with reliable acceleration profiles reached their maximum acceleration below 2~$R_\odot$ and 38\% of CMEs reached their maximum velocity within 2~$R_\odot$. \citet{bein2011impulsive} studied the evolution of 94 CMEs using STEREO EUVI and COR instruments and found that 96\% of events reached their maximum acceleration below 2~$R_\odot$ and 81\% reached their maximum velocity below 2~$R_\odot$.
These results suggest that the dominant effect in the decrease in correlation comes from the coronagraphs missing the early acceleration phase of the CME rather than the projection effects.

In summary, we observe a positive weak correlation between the maximum CME velocity $v_\text{max}$ and characteristic dimming parameters. However, we find that this correlation weakens significantly when CME properties are derived solely from coronagraph observations rather than incorporating EUV imaging data.

\section{Summary and discussion}
We present an analysis on coronal dimmings associated with flare and CME activity from a single active region, AR~13664, during the first half of May 2024 using SDO/AIA and HMI data. 
This AR exhibited remarkable activity, producing 12 X-class flares and 55 M-class flares during its first rotation on the solar disc, along with multiple CMEs that ultimately triggered the strongest geomagnetic storm in two decades. We analyse 67 distinct flare events $\geq$M1.0 class between May 1 and May 15, and identified associated coronal dimmings and CMEs. While both on-disc and off-limb dimmings are identified, our study primarily focuses on the analysis of on-disc events. The key findings of our analysis are summarised as follows:
\begin{enumerate}
    \item Coronal dimmings are detected in 22 events, of which 16 are observed on-disc and 6  are observed off-limb over the western limb. Among the on-disc dimmings six are associated with X-class flares and ten to M-class flares. Off-limb dimmings were linked to four X-class and two M-class flares (see Table~\ref{table:flare_list}, last column).
    \item A total of 23 CMEs are associated with the flares from AR~13664, with ten linked to X-class flares, and 13 to M-class flares (see Table~\ref{table:flare_list}, `CME' column). Thus, 83\% of X-class flares are accompanied by a CME, compared to 23\% of M-class flares. Among these CMEs, ten were Earth-directed halo CMEs. 
    \item Among the 16 on-disc dimmings, 13 are associated with CMEs, while all 6 off-limb dimmings had corresponding CMEs. This means that 3 dimmings observed on-disc lacked an identifiable CME counterpart. On the other hand, 4 CMEs revealed no detected coronal dimming.
    \item The average area of the 16 on-disc dimmings under study is $(3.42\pm2.05)\times10^{10}$~km$^2$ and contains total unsigned magnetic flux exceeding $5.5\times10^{21}$~Mx. The mean unsigned magnetic flux density underlying the segmented dimming regions projected to the solar photosphere is on average $(121.12\pm73.13)$~G. 
    \item The average brightness drop in the dimming regions is $(-1.68\pm0.55)\times10^6$~DN, corresponding to a mean relative intensity decrease of $61$\% with respect to the pre-eruption levels.
    \item The dimming area parameters ($A$, $\dot{A}$, $A_\phi$, $\dot{A}_\phi$) exhibit a strong correlation with the  GOES SXR peak flux $F_P$ and fluence $F_T$, with the strongest correlation observed between the magnetic dimming area $A_\phi$ and $F_P$, $c=0.85\pm0.08$. The total unsigned magnetic dimming flux $\phi$ shows a moderate correlation with the GOES SXR peak flux of the associated flare $F_P$ ($c=0.58\pm0.14$) and the SXR flare fluence $F_T$ ($c=0.69\pm0.19$).
    \item No significant correlation is found between dimming characteristic parameters and the peak of the time derivative of the flare SXR flux $\dot{F}_P$. Additionally, the mean brightness drop $\bar{I}_\text{drop}$ and the mean magnetic flux density $B_\text{us}$ in the dimming region exhibit anticorrelations with the flare peak SXR flux $F_P$ of $c=-0.41\pm0.24$ and $c=-0.57\pm0.25$, respectively. This indicates that coronal dimmings and flares are mostly related via the dimming area and growth rate when focusing on a single AR.
    \item The maximum CME velocity derived from SOHO/LASCO measurements correlates well with area-dependent dimming parameters, such as $\dot{A}$ ($c=0.68\pm0.16$), $\dot{I}_\text{drop}$  ($c=0.67\pm0.11$) and $\phi$ ($c=0.70\pm0.10$). However, no significant correlation is observed with area-independent parameters, which are the mean brightness drop $\bar{I}_\text{drop}$ ($c=-0.03\pm0.28$) and the mean magnetic flux density $B_\text{us}$ ($c=-0.34\pm0.29$).
\end{enumerate}

For the first time, coronal dimmings originating from the same AR have been continuously monitored over a 14-day period. Most previous dimming studies have focused on individual events or a collection of unrelated events. An exception is the work by \citet{mason2016relationship}, who studied dimmings detected in SDO/EVE irradiance observations over two separate two-week periods. However, that study was focused on spatially integrated data, detaching the dimming analysis from the ARs of origin. In contrast, the present analysis provides a unique opportunity to gain deeper insights into the relationship between coronal dimmings and the ARs in which they arise, as well as their temporal evolution.

\subsection{Relation of AR~13664 dimming parameters to general dimming population}
The magnetic properties of the dimmings in this study evidence the large magnetic flux density of AR~13664. Both the total dimming magnetic flux $\phi$ and mean magnetic flux density $B_\text{us}$ significantly exceed the average values reported in \citet{Dissauer2018b}. For example, in the May 2024 events $B_\text{us}$ is $(121.12\pm73.13)$~G compared to $(74.90\pm47.62)$~G in \citet{Dissauer2018b}. AR~13664 is also among the largest ARs in recent decades \citep{hayakawa2024solar}, which is also reflected in the dimming areas $A$ being larger for a given area growth rate $\dot{A}$ compared to \citet{Dissauer2018b}. 

Furthermore, we find an increased correlation between the GOES SXR peak flux $F_P$, fluence $F_T$, and various dimming parameters -- particularly area-related parameters like total dimming area $A$ and area growth rate $\dot{A}$ -- compared to the correlations reported by \citet{Dissauer2018b}. This enhancement may be expected, as all events analysed occur within the same AR, and may thus share some basic characteristics. Consequently, variations in energy release directly influence the associated dimmings without introducing additional uncertainties related to changes in AR size and morphology. Furthermore, \citet{Dissauer2018b} noted that stronger flares ($\geq$M1.0) exhibit a better correlation between dimming area and flare ribbon area than weaker flares. Given that the ribbon area is itself also correlated with the peak flare flux \citep{Kazachenko2017database}, the fact that our study exclusively considers M- and X-class flares may further contribute to the increased correlation observed in our study.

Additionally, we find a notable increase in the anticorrelation between the mean magnetic flux density $B_\text{us}$, the mean intensity drop $\bar{I}_\text{drop}$, and the dimming area parameters, a relationship that was absent in the analysis of \citet{Dissauer2018b}. These parameters are normalised by dimming area. Thus, in the case of AR~13664, the underlying magnetic configuration remains relatively unchanged across different dimming events, so $B_\text{us}$ decreases as the dimming expands into weaker magnetic field regions. Similarly, intensity drops are more pronounced near the core of the AR with strong coronal loops but weaken as the dimming extends outwards, so a large dimming expansion into surrounding regions leads to a lower $\bar{I}_\text{drop}$. However, these anticorrelations do not contradict \citet{Dissauer2018b}, as they are not evident when considering the entire combined dataset. But studying many events from one single flare-productive active region like AR~13664 provides a means to identify relations to the magnetic properties of the host AR. Such findings may be also relevant when extending the statistical relations towards other stars, where dimmings may provide an important means to detect stellar CMEs \citep{Veronig2021indications, Loyd2022stellar}.

\subsection{Relation of dimmings to CMEs and eruptive versus confined flares}

Alongside the dimming detection, we also studied the occurrence of major flares ($\geq$M1.0) and CMEs originating from AR~13664, which are documented in Table~\ref{table:flare_list}. This AR is among the most flare-productive in the GOES observation history, particularly for M- and X-class flares. The majority of these flares were confined, with 77\% of M-class flares and 17\% of X-class flares not being associated with a CME. These findings contrast with general statistical studies, which indicate that approximately 40\% of M-class flares and 10\% of X-class flares are typically confined \citep{Yashiro2006different}. 

The excess of the number of M- and X-class flares that are confined in the May 2024 events is related to the magnetic confinement of the source AR. As \citet{Li2021magnetic} have shown in their statistics, large flares that are produced by ARs with a large amount of magnetic flux are more likely to be confined, due to the strong overlying fields in these ARs. AR~13664 is such an exemplary case. Specifically, as reported in \citep{Li2021magnetic},when the magnetic flux exceeds $1.0\times10^{23}$~Mx, approximately 97\% of the M- and X-class flares are not associated with a CME. The maximum total unsigned magnetic flux and magnetic free energy of AR~13664 are comparable to those of AR NOAA 12192 in October 2014 \citep{Jarolim2024magnetic}, which exhibited the largest unsigned magnetic flux and magnetic free energy during Solar Cycle 24 \citep{Sun2015why, Karimov20243d}, and was very unique in producing large confined flares \citep{Thalmann2015confined}. In total, AR~12192 caused 6 X-class and 29 M-class flares -- which were all confined except one \citep{Veronig2015magnetic}. The large magnetic flux values of AR~13664, of up to $1.9\times10^{23}$~Mx for the unsigned magnetic flux \citep{Jarolim2024magnetic, Romano2024analyzing, hayakawa2024solar}, may explain the low eruptivity of the energetic flares observed in AR~13664, with the strong magnetic fields inhibiting a successful eruption.

Several cases were identified where expected dimmings were absent. Notably, two CMEs early in AR~13664's development (May 2-3) showed no clear dimming signatures. This absence is likely due to a combination of projection effects and the weaker nature of the associated M2.7-class flares, which may have resulted in smaller-scale plasma evacuations that were more difficult to detect. Another case is the M5.9 class flare on May 10, which was linked to a CME but lacked a distinct dimming. This event coincided with significant magnetic restructuring in the southern hemisphere outside AR~13664, suggesting that the CME may have originated from a broader reconfiguration rather than a localised eruption in the region.

On the other hand, three dimmings were detected without an associated CME (see Table~\ref{table:flare_list}).  All were linked to M-class flares, and two (May 5 at 18:34 UT and May 9 at 06:03 UT) had slow area growth rates, peaking about an hour after flare onset at values below $5\times10^6$ km$^2$~s$^{-1}$. \citet{Veronig2021indications} found that, based on spatially integrated SDO/EVE and AIA data, the conditional probability of detecting a dimming without a CME was 0.125 for flares above M5-class. Given the 48 confined flares in our dataset, these three detections align with the expectations. However, as most studies focus on CME-associated dimmings, further research on the role of dimmings in confined flares is needed.

A key observation in our study is the significant difference in correlation between dimming properties and maximum CME velocity when derived from SOHO/LASCO measurements compared to STEREO/EUVI+COR measurements. Previous studies have shown a connection between dimmings and CME kinematics \citep{mason2016relationship, Dissauer2019, chikunova2020coronal}. Specifically, \citet{Dissauer2019} reported a strong correlation between second-order dimming parameters; that is, the change rate of the dimming region, and the maximum velocity reached by the CMEs. While a similar correlation appears on the May 2024 events under study, we note that the correlations reported by \citet{Dissauer2019} would have been significantly weaker if only coronagraphic measurements (from either STEREO or LASCO) had been utilised. This discrepancy arises because the combined use of the STEREO/EUVI+COR instruments tracks the CME from its source in the lower corona onwards, whereas LASCO first detects CMEs at approximately 2~$R_\odot$. As a consequence, the main acceleration phase of the CME, which has been shown to be closely associated with the impulsive phase of the dimming \citep{miklenic2011coronal, Dissauer2019}, is not captured in coronagraphic data. This may explain why some previous studies encountered challenges in identifying a robust correlation between CME speeds and coronal dimming parameters when relying exclusively on coronagraph observations \citep{reinard2009relationship, aschwanden2016global, krista2017statistical, Vakhrusheva2024parameters}.

\subsection{Problematics of dimming detections and identifications}

The criteria by which coronal dimmings are identified presents a specially great challenge when continuously analysing coronal dimmings in time. Most statistical studies rely on visual preselection to exclude unrelated processes, such as plasma heating or cooling beyond the detector's passband or the removal of bright features from the observation frame, from being misclassified as dimmings \citep{reinard2009relationship, mason2016relationship, krista2017statistical, krista2022study, Dissauer2018b, Dissauer2019}. Given the strong dynamics of AR~13664, `false dimmings' -- emission decreases unrelated to a mass depletion -- are frequently observed throughout the analysed time range. Consequently, a preselection step was necessary for this study. In general, we note that the big size and the large coronal brightness of AR~13664 (see, e.g. Fig.~\ref{fig:overview-example}) together with the high frequency of events occurring in sequence makes the detection of dimming regions more difficult than in other ARs.

\citet{attrill2010automatic} suggested that the duration of a detected event serves as a primary criterion to differentiate genuine coronal dimmings from spurious events, since real dimmings persist for extended periods of time \citep{attrill2008recovery, vanninathan2018plasma, ronca2024recovery}. However, the frequent flare activity of AR~13664 discouraged us from any long-term analysis, as multiple flares often occurred within an hour. To address this, we performed a superposed epoch analysis to determine which characteristic parameter evolution best discriminates real dimmings from false detections. The dimming area $A(t)$ and the area growth rate $\dot{A}(t)$ exhibited the most distinct evolution patterns in terms of real and false dimmings (see Fig.~\ref{fig:selection}). Based on these findings, we established a minimum threshold for the mean dimming area growth rate of $1.8\times10^6$~km$^2$~s$^{-1}$ within the first hour following flare onset to qualify as a real dimming. Using this criterion for on-disc dimmings, all CMEs associated with $\geq$M1.0 class flares reported by \citet{hayakawa2024solar} were successfully matched to a corresponding dimming event. On the downside, this threshold prevents us from observing dimmings of a smaller extent than $6.5\times10^9$~km$^2$.

\section{Conclusions}
This study presents the first systematic analysis of spatially resolved mass-loss coronal dimmings and their relation to flares and CMEs within a single active region over a two-week period, providing new insights into the relation between the AR properties and the dimmings it produces. We find that dimmings associated with AR~13664 exhibit a stronger correlation with their associated flares compared to the general population of previously studied dimmings. Additionally, the dimming magnetic flux is significantly larger, reflecting the exceptionally strong magnetic flux and energy of the AR. The strong magnetic fields of AR~13664 also explain the high occurrence of confined M- and X-class flares, exceeding the average flare-CME association rate, which we attribute to the enhanced magnetic confinement in large ARs. This finding is particularly relevant for stellar CME studies, where strong magnetic fields may contribute to the low detection rate of stellar CMEs. Furthermore, we confirm that dimming evolution reflects CME dynamics in the low corona, reinforcing the role of dimmings as a key diagnostic for early CME acceleration.
Our findings further emphasize the extensive potential of coronal dimmings in improving the understanding and characterisation of solar eruptions. 

\begin{acknowledgements}
This project has received funding from the European Union's Horizon Europe research and innovation programme under grant agreement No 101134999 (SOLER). The research was sponsored by the DynaSun project and has thus received funding under the Horizon Europe programme of the European Union under grant agreement (no. 101131534). Views and opinions expressed are however those of the author(s) only and do not necessarily reflect those of the European Union and therefore the European Union cannot be held responsible for them. SDO data are courtesy of NASA/SDO and the AIA, and HMI science teams. GOES is a joint effort of NASA and the National Oceanic and Atmospheric Administration (NOAA). The CDAW CME catalogue is generated and maintained at the CDAW Data Center by NASA and The Catholic University of America in cooperation with the Naval Research Laboratory. SOHO is a project of international cooperation between ESA and NASA.
\end{acknowledgements}

\bibliographystyle{aa}
\bibliography{aa54772-25}

\begin{appendix}
    \onecolumn
    \section{May 2024 flare event list and flare-CME-dimming associations}\label{Appendix_tables}
    We identified all M- and X-class flares, CMEs, and coronal dimmings originating from AR~13664 between May 1 and 15, following the methodology described in Sect.\ref{sec:data:selection}, and established associations between these events. The complete list of identified flares, along with their corresponding CMEs and coronal dimmings, is provided in Table~\ref{table:flare_list}.
    
    \begin{longtable}{cccccccccccc}
    \caption{Flare list (M- and X-class) of AR~13664 from May 1 to 15, 2024}\\
    \hline\hline
    N & Day & Start & Peak & End & \CellWithForceBreak{Flare \\ Location} & \CellWithForceBreak{F$_P$ \\ (W m$^{-2}$)} & \CellWithForceBreak{$\dot{F}_P$ \\ (W m$^{-2}$s$^{-1}$)} & \CellWithForceBreak{F$_T$ \\ (J m$^{-2}$)} & CME &  \CellWithForceBreak{$v_\text{max}$\\ (km~s$^{-1}$) } & Dim \\ 
    \hline
    \endfirsthead
    \caption{(continued)}\\
    \hline\hline
    N & Day & Start & Peak & End & \CellWithForceBreak{Flare \\ Location} & \CellWithForceBreak{F$_P$ \\ (W m$^{-2}$)} & \CellWithForceBreak{$\dot{F}_P$ \\ (W m$^{-2}$s$^{-1}$)} & \CellWithForceBreak{F$_T$ \\ (J m$^{-2}$)} & CME &  \CellWithForceBreak{$v_\text{max}$\\ (km~s$^{-1}$)} & Dim \\
    \hline
    \endhead
    \hline
    \endfoot
    
    1 & 2 & 20:52 & 20:57 & 21:01 & S18 E58 & 2.86E-05 & 5.45E-07 & 6.61E-03 & 02-214916* & 142.2 & --- \\
    2 & 3 & 00:08 & 00:15 & 00:19 & S19 E56 & 1.09E-05 & 2.17E-06 & 7.94E-04 & 03-000005* & 187.6 & --- \\
    3 & 4 & 18:10 & 18:20 & 18:25 & S16 E29 & 1.33E-05 & 1.29E-07 & 6.55E-03 & --- & --- & --- \\
    4 & 5 & 09:23 & 09:38 & 09:53 & S18 E23 & 2.37E-05 & 7.01E-08 & 3.34E-02 & --- & --- & --- \\
    5 & 5 & 14:33 & 14:47 & 15:19 & S20 E16 & 1.31E-05 & 3.00E-08 & 2.46E-02 & 05-153607 & 623.5   & DISC \\
    6 & 5 & 16:55 & 17:01 & 17:06 & S20 E16 & 1.38E-05 & 4.63E-08 & 7.13E-03 & --- & --- & --- \\
    7 & 5 & 18:34 & 18:40 & 18:45 & S17 E21 & 1.07E-05 & 8.30E-08 & 5.16E-03 & --- & --- & DISC \\
    8 & 6 & 09:49 & 09:59 & 10:04 & S16 E13 & 1.57E-05 & 5.90E-08 & 8.48E-03 & --- & --- & --- \\
    9 & 7 & 00:41 & 00:58 & 01:23 & S18 E06 & 2.67E-05 & 3.28E-08 & 5.46E-02 & 07-023605* & 483.6 & DISC \\
    10 & 7 & 08:18 & 08:23 & 08:40 & S18 E01 & 1.50E-05 & 1.01E-07 & 1.44E-02 & --- & --- & --- \\
    11 & 7 & 11:40 & 11:50 & 12:01 & S19 E00 & 2.49E-05 & 1.56E-07 & 2.09E-02 & 07-154807* & 444.6 & DISC \\
    12 & 7 & 13:32 & 13:35 & 13:39 & S18 W05 & 1.06E-05 & 4.83E-07 & 4.25E-03 & --- & --- & --- \\
    13 & 7 & 19:58 & 20:22 & 20:34 & S18 W09 & 2.17E-05 & 2.21E-07 & 2.56E-02 & --- & --- & --- \\
    14 & 8 & 02:16 & 02:27 & 02:36 & S17 W04 & 3.45E-05 & 1.07E-07 & 2.78E-02 & 08-023607 & 812.4 & --- \\
    15 & 8 & 03:19 & 03:27 & 03:38 & S18 W06 & 1.91E-05 & 8.82E-08 & 1.81E-02 & 08-023607* & 812.4 & DISC \\
    16 & 8 & 04:37 & 05:09 & 05:32 & S19 W07 & 1.04E-04 & 1.43E-07 & 2.39E-01 & 08-053606$^\dagger$ & 1102.5 & DISC \\
    17 & 8 & 11:13 & 12:04 & 12:36 & S17 W08 & 8.60E-05 & 2.21E-07 & 2.54E-01 & 08-122405$^\dagger$ & 1321.1 & DISC \\
    18$^\wedge$ & 8 & 17:32 & 17:53 & 18:00 & S19 W18 & 7.95E-05 & 3.17E-07 & 5.63E-02 & --- & --- & --- \\
    19$^\wedge$ & 8 & 19:15 & 19:21 & 19:29 & S19 W19 & 2.08E-05 & 1.02E-07 & 1.53E-02 & --- & --- & DISC \\
    21$^\wedge$ & 8 & 21:12 & 21:40 & 23:19 & S19 W21 & 1.00E-04 & 1.60E-07 & 5.47E-01 & 08-222405$^\dagger$ & 1304.4 & DISC \\
    22 & 9 & 03:07 & 03:17 & 03:23 & S20 W23 & 4.11E-05 & 3.42E-07 & 2.05E-02 & --- & --- & --- \\
    23 & 9 & 03:23 & 03:32 & 03:45 & S20 W23 & 4.59E-05 & 1.27E-07 & 5.08E-02 & --- & --- & --- \\
    24 & 9 & 04:44 & 04:49 & 04:55 & S19 W25 & 1.76E-05 & 1.57E-07 & 8.79E-03 & --- & --- & --- \\
    25 & 9 & 06:03 & 06:13 & 06:24 & S20 W24 & 2.35E-05 & 9.72E-08 & 2.31E-02 & --- & --- & DISC \\
    26 & 9 & 08:45 & 09:13 & 09:36 & S20 W23 & 2.24E-04 & 3.77E-07 & 4.34E-01 & 09-092405$^\dagger$ & 1962.8 & DISC \\
    27 & \CellWithForceBreak{9 \\ 9} & \CellWithForceBreak{11:52 \\ 12:05} & \CellWithForceBreak{11:56 \\ 12:12} & \CellWithForceBreak{12:02 \\ 12:20} & \CellWithForceBreak{S17 W34 \\ S17 W31} & \CellWithForceBreak{3.19E-05 \\ 3.01E-05} & \CellWithForceBreak{2.98E-07 \\ 1.32E-07} & \CellWithForceBreak{1.34E-02 \\ 2.30E-02} & \CellWithForceBreak{09-122406} & \CellWithForceBreak{1035.5 } & \CellWithForceBreak{ DISC }  \\
    28 & 9 & 13:16 & 13:23 & 13:29 & S17 W29 & 3.79E-05 & 3.90E-07 & 1.80E-02 & --- & --- & --- \\
    29 & 9 & 17:23 & 17:44 & 18:00 & S17 W28 & 1.12E-04 & 2.22E-05 & 1.40E-01 & 09-185200$^\dagger$ & 1454.0 & DISC \\
    30 & 9 & 23:04 & 23:08 & 23:13 & S18 W37 & 1.27E-05 & 5.59E-08 & 6.07E-03 & --- & --- & --- \\
    31 & 9 & 23:44 & 23:51 & 23:55 & S18 W37 & 1.63E-05 & 8.61E-08 & 6.93E-03 & --- & --- & --- \\
    32 & 10 & 00:10 & 00:13 & 00:22 & S19 W34 & 1.51E-05 & 1.59E-07 & 9.01E-03 & --- & --- & --- \\
    33 & 10 & 03:15 & 03:29 & 03:40 & S19 W35 & 1.46E-05 & 3.07E-08 & 1.69E-02 & --- & --- & --- \\
    34 & 10 & 06:27 & 06:54 & 07:06 & S17 W34 & 3.95E-04 & 9.38E-07 & 4.32E-01 & 10-071205$^\dagger$ & 1249.5 & DISC \\
    35 & 10 & 10:10 & 10:14 & 10:19 & S18 W36 & 2.31E-05 & 1.60E-07 & 9.26E-03 & --- & --- & --- \\
    36 & 10 & 13:58 & 14:11 & 14:23 & S17 W39 & 5.99E-05 & 1.94E-07 & 5.49E-02 & 10-162405* & 412.7 & ---    \\
    37 & 10 & 18:26 & 18:32 & 18:38 & S17 W44 & 1.17E-05 & 2.08E-06 & 6.01E-03 & --- & --- & --- \\
    38 & 10 & 18:38 & 18:48 & 18:57 & S17 W43 & 1.81E-05 & 3.35E-06 & 1.57E-02 & --- & --- & --- \\
    39 & 10 & 18:57 & 19:05 & 19:10 & S17 W42 & 2.03E-05 & 6.22E-08 & 1.43E-02 & --- & --- & --- \\
    40 & 10 & 19:35 & 19:53 & 19:56 & S16 W40 & 1.19E-05 & 2.88E-08 & 1.28E-02 & --- & --- & --- \\
    41 & 10 & 19:56 & 20:03 & 20:18 & S17 W41 & 1.96E-05 & 6.23E-08 & 2.16E-02 & --- & --- & --- \\
    42 & 10 & 20:59 & 21:08 & 21:12 & S18 W46 & 3.94E-05 & 5.08E-07 & 1.68E-02 & --- & --- & --- \\
    43 & 11 & 01:10 & 01:23 & 01:39 & S17 W44 & 5.88E-04 & 2.45E-06 & 6.21E-01 & 11-013605$^\dagger$ & 1981.7 & DISC \\
    44 & 11 & 10:03 & 10:18 & 10:34 & S19 W53 & 3.13E-05 & 6.43E-08 & 4.48E-02 & --- & --- & --- \\
    45 & 11 & 10:53 & 10:56 & 11:00 & S19 W54 & 1.77E-05 & 1.11E-07 & 5.56E-03 & --- & --- & --- \\
    46 & 11 & 11:15 & 11:44 & 12:05 & S18 W63 & 1.53E-04 & 3.29E-07 & 2.59E-01 & --- & --- & --- \\
    47 & 11 & 13:45 & 13:49 & 14:08 & S18 W63 & 1.85E-05 & 1.12E-07 & 2.06E-02 & --- & --- & --- \\
    48 & 11 & 14:46 & 15:25 & 15:52 & S19 W64 & 8.93E-05 & 1.72E-07 & 1.97E-01 & 11-161205 & 1058.0 & DISC \\
    49 & 11 & 20:32 & 20:41 & 20:47 & S19 W65 & 1.26E-05 & 7.23E-08 & 7.60E-03 & --- & --- & --- \\
    50 & 12 & 00:41 & 00:45 & 00:52 & S19 W66 & 3.27E-05 & 4.16E-07 & 1.53E-02 & --- & --- & --- \\
    51 & 12 & 05:37 & 05:52 & 06:06 & S17 W64 & 2.43E-05 & 5.93E-08 & 2.85E-02 & --- & --- & --- \\
    52 & 12 & 12:27 & 12:41 & 12:54 & S17 W66 & 1.68E-05 & 5.22E-08 & 1.88E-02 & --- & --- & --- \\
    53 & 12 & 13:49 & 13:56 & 14:08 & S19 W71 & 1.57E-05 & 4.57E-08 & 1.50E-02 & --- & --- & --- \\
    54 & 12 & 16:11 & 16:26 & 16:38 & S19 W72 & 1.02E-04 & 4.27E-07 & 9.46E-02 & --- & --- & --- \\
    55 & 12 & 20:17 & 20:32 & 20:49 & S17 W73 & 4.97E-05 & 2.08E-07 & 5.56E-02 & --- & --- & --- \\
    56 & 12 & 22:01 & 22:06 & 22:12 & S17 W74 & 1.17E-05 & 6.73E-08 & 6.25E-03 & --- & --- & --- \\
    57 & 12 & 23:00 & 23:10 & 23:14 & S17 W75 & 1.06E-05 & 1.12E-07 & 5.23E-03 & --- & --- & --- \\
    58 & 13 & 01:23 & 01:33 & 01:38 & S19 W75 & 1.39E-05 & 1.44E-07 & 8.34E-03 & --- & --- & --- \\
    59 & 13 & \CellWithForceBreak{08:06 \\ 08:23 \\ 08:48} & \CellWithForceBreak{08:20 \\ 08:29 \\ 09:44} & \CellWithForceBreak{08:23 \\ 08:33 \\ 10:57} & \CellWithForceBreak{S19 W79 \\ S19 W80 \\ S19 W77} & \CellWithForceBreak{1.31E-05 \\ 1.47E-05 \\ 6.72E-05} & \CellWithForceBreak{6.30E-08 \\ 2.56E-08 \\ 3.39E-07} & \CellWithForceBreak{9.51E-03 \\ 8.21E-03 \\ 3.28E-01} & 13-084806$^\dagger$ & 2603.5 & LIMB \\
    60 & 13 & 12:56 & 13:11 & 13:23 & S18 W80 & 3.73E-05 & 9.65E-08 & 4.53E-02 & --- & --- & --- \\
    61 & 13 & 21:48 & 21:59 & 22:07 & S18 W85 & 1.61E-05 & 7.97E-08 & 1.15E-02 & --- & --- & --- \\
    62 & 14 & 01:23 & 01:48 & 02:04 & S18 W83 & 2.56E-05 & 4.54E-08 & 4.11E-02 & 14-013605 & 613.1 &     \\
    63 & 14 & 02:03 & 02:09 & 02:19 & S17 W88 & 1.71E-04 & 1.78E-06 & 1.09E-01 & 14-020005 & 1010.7 & LIMB \\
    64 & 14 & 12:40 & 12:55 & 13:05 & S17 W78 & 1.24E-04 & 5.62E-07 & 9.94E-02 & 14-130005  & 1307.6 & LIMB \\
    65 & 14 & 16:46 & 16:51 & 17:02 & S17 W89 & 8.58E-04 & 6.15E-06 & 5.35E-01 & 14-170005$^\dagger$ & 2880.5 & LIMB \\
    66 & 15 & 06:23 & 08:37 & 08:51 & S18 W83 & 3.38E-04 & 6.15E-06 & 5.35E-01 & 15-083607$^\dagger$ & 2016.8 & LIMB \\
    67 & 15 & 10:00 & 10:35 & 11:57 & WL & 3.55E-05 & 4.46E-08 & 1.96E-01 & 15-104805 & 1472.2  & LIMB     
    \label{table:flare_list}\end{longtable}
    \tablefoot{We list the day, start time and end time of the associated flare derived from the GOES flare catalogue, or directly derived from the GOES $1.0 - 8.0$~\AA~SXR flux; the peak of the GOES $1.0 - 8.0$~\AA~SXR flux $F_P$; the maximum of its time derivative $\dot{F}_P$; the SXR flare fluence $F_T$; and the location of the flare. The CME column shows the ID of the associated CME as found in the SOHO/LASCO catalogue. Halo CMEs are marked with a $(^\dagger)$, and CMEs that do not appear in \citet{hayakawa2024solar} are marked with an asterisk (*). The maximum velocity of the CMEs ($v_\text{max}$) derived from LASCO height-time measurements is also given. The last column (`Dim') indicates whether a dimming is found in connection with the flare and if it is seen on-disc (DISC) or off-limb (LIMB). Events with $^\wedge$ required the alternative data processing.}
    \FloatBarrier

    \begin{sidewaystable}
    \section{Characteristic dimming parameters for the May 2024 events dimmings}
        The characteristic dimming parameters described in Sect.~\ref{sec:methods:parameters} detected on-disc for the May 2024 events dimmings using the coronal dimming detection method described in Sect.~\ref{sec:methods:detection} are provided in Table~\ref{table:dimming_properties}. 
    
    \caption{\label{table:dimming_properties}Characteristic dimming properties of the on-disc coronal dimmings from AR~13664}   
    \centering
    \begin{tabular}{cccccccccccc}         
    \hline\hline
    N & Day & \CellWithForceBreak{Flare\\Start} & \CellWithForceBreak{$A$\\ ($10^{10}$~km$^2$)} &  \CellWithForceBreak{$\dot{A}$\\ ($10^{7}$~km$^2$ s$^{-1}$)} &  \CellWithForceBreak{$\phi$\\ ($10^{21}$~Mx)} &  \CellWithForceBreak{$\dot{\phi}$\\ ($10^{18}$~Mx~s$^{-1}$)} &  \CellWithForceBreak{$\phi_+$ \\ ($10^{21}$~Mx)} &  \CellWithForceBreak{$|\phi_-|$\\ ($10^{21}$~Mx)} &  \CellWithForceBreak{$B_{\text{us}}$\\ (G)} &  \CellWithForceBreak{$I_{\text{drop}}$\\ ($10^5$~DN)} &  \CellWithForceBreak{$\dot{I}_{\text{drop}}$\\ ($10^2$~DN~s$^{-1}$)}  \\
    \hline         
    4 & 5 & 14:33 & $ 2.91\pm 0.36 $ & $ 0.84\pm 0.07 $ & $ 5.57\pm 0.56 $ & $ 2.60\pm 0.04 $ & $ 3.01\pm 0.15 $ & $ 2.56\pm 0.41 $ & $ 67.97\pm 0.07 $ & $ -5.99\pm 0.26 $ & $ -3.42\pm 0.01 $ \\
    6 & 5 & 18:34 & $ 1.89\pm 0.27 $ & $ 0.48\pm 0.07 $ & $ 5.55\pm 0.58 $ & $ 3.16\pm 0.05 $ & $ 3.11\pm 0.28 $ & $ 2.44\pm 0.30 $ & $ 86.42\pm 2.17 $ & $ -18.60\pm 1.25 $ & $ -6.01\pm 0.29 $ \\
    8 & 7 & 00:41 & $ 2.81\pm 0.35 $ & $ 0.61\pm 0.09 $ & $ 8.29\pm 1.02 $ & $ 5.44\pm 0.88 $ & $ 3.40\pm 0.43 $ & $ 4.89\pm 0.59 $ & $ 110.85\pm 0.97 $ & $ -8.77\pm 0.34 $ & $ -5.78\pm 1.16 $ \\
    10 & 7 & 11:40 & $ 1.67\pm 0.21 $ & $ 0.50\pm 0.01 $ & $ 8.49\pm 1.95 $ & $ 3.77\pm 0.08 $ & $ 3.29\pm 0.71 $ & $ 5.21\pm 1.24 $ & $ 156.29\pm 6.88 $ & $ -16.05\pm 1.86 $ & $ -6.71\pm 1.29 $ \\
    14 & 8 & 03:19 & $ 3.03\pm 0.42 $ & $ 1.57\pm 0.21 $ & $ 8.99\pm 1.33 $ & $ 3.56\pm 0.05 $ & $ 4.66\pm 0.62 $ & $ 4.33\pm 0.72 $ & $ 109.63\pm 1.28 $ & $ -14.39\pm 0.30 $ & $ -8.58\pm 0.22 $ \\
    15 & 8 & 04:37 & $ 5.20\pm 0.60 $ & $ 1.49\pm 0.08 $ & $ 14.74\pm 1.71 $ & $ 6.49\pm 0.28 $ & $ 4.97\pm 0.87 $ & $ 9.78\pm 0.84 $ & $ 100.82\pm 0.14 $ & $ -26.26\pm 1.27 $ & $ -25.37\pm 1.59 $ \\
    16 & 8 & 11:13 & $ 2.70\pm 0.28 $ & $ 0.90\pm 0.08 $ & $ 9.82\pm 1.47 $ & $ 3.43\pm 0.09 $ & $ 6.60\pm 0.80 $ & $ 3.22\pm 0.67 $ & $ 92.42\pm 0.60 $ & $ -22.73\pm 0.18 $ & $ -11.24\pm 0.01 $ \\
    19 & 8 & 19:15 & $ 1.69\pm 0.26 $ & $ 1.22\pm 0.13 $ & $ 20.28\pm 2.94 $ & $ 10.45\pm 0.74 $ & $ 10.32\pm 1.20 $ & $ 9.96\pm 1.74 $ & $ 352.33\pm 4.45 $ & $ -17.20\pm 1.20 $ & $ -12.36\pm 1.85 $ \\
    21 & 8 & 21:12 & $ 3.53\pm 0.51 $ & $ 1.29\pm 0.06 $ & $ 26.23\pm 3.06 $ & $ 10.50\pm 0.94 $ & $ 10.11\pm 1.37 $ & $ 16.12\pm 1.69 $ & $ 205.16\pm 6.96 $ & $ -18.87\pm 1.39 $ & $ -13.68\pm 0.61 $ \\
    27 & 9 & 06:03 & $ 1.16\pm 0.14 $ & $ 0.31\pm 0.02 $ & $ 5.68\pm 0.99 $ & $ 4.42\pm 0.84 $ & $ 2.46\pm 0.38 $ & $ 3.22\pm 0.60 $ & $ 167.68\pm 2.39 $ & $ -11.35\pm 0.76 $ & $ -6.87\pm 0.61 $ \\
    28 & 9 & 08:45 & $ 2.64\pm 0.40 $ & $ 1.14\pm 0.15 $ & $ 17.09\pm 2.04 $ & $ 5.53\pm 0.85 $ & $ 4.39\pm 0.76 $ & $ 12.71\pm 1.28 $ & $ 136.58\pm 4.04 $ & $ -20.28\pm 0.07 $ & $ -19.23\pm 0.89 $ \\
    29 & 9 & 11:50 & $ 1.45\pm 0.23 $ & $ 0.68\pm 0.08 $ & $ 6.63\pm 1.90 $ & $ 2.66\pm 0.96 $ & $ 1.27\pm 0.20 $ & $ 5.36\pm 1.71 $ & $ 110.94\pm 2.26 $ & $ -24.74\pm 2.11 $ & $ -9.59\pm 0.00 $ \\
    33 & 9 & 17:23 & $ 5.58\pm 0.68 $ & $ 2.35\pm 0.25 $ & $ 16.54\pm 4.27 $ & $ 8.87\pm 2.42 $ & $ 5.99\pm 2.00 $ & $ 10.55\pm 2.26 $ & $ 58.15\pm 4.05 $ & $ -23.39\pm 1.93 $ & $ -14.89\pm 0.54 $ \\
    38 & 10 & 06:27 & $ 7.27\pm 0.74 $ & $ 3.66\pm 0.56 $ & $ 13.66\pm 2.96 $ & $ 7.61\pm 2.67 $ & $ 4.84\pm 1.42 $ & $ 8.81\pm 1.55 $ & $ 57.05\pm 3.33 $ & $ -23.27\pm 0.73 $ & $ -21.07\pm 0.93 $ \\
    47 & 11 & 01:10 & $ 8.37\pm 0.92 $ & $ 5.99\pm 0.93 $ & $ 17.77\pm 2.34 $ & $ 10.53\pm 0.31 $ & $ 5.52\pm 0.70 $ & $ 12.25\pm 1.64 $ & $ 32.98\pm 0.17 $ & $ -21.95\pm 0.83 $ & $ -15.09\pm 0.28 $ \\
    52 & 11 & 14:46 & $ 2.77\pm 0.28 $ & $ 1.61\pm 0.09 $ & $ 15.30\pm 4.91 $ & $ 10.06\pm 3.12 $ & $ 7.91\pm 0.83 $ & $ 7.39\pm 4.08 $ & $ 124.62\pm 19.65 $ & $ -15.91\pm 0.55 $ & $ -10.29\pm 0.26 $ \\
    \hline 
    \end{tabular}
    \tablefoot{We list a selection of the dimming parameters for which the number in the N column refers to the numbering in Table~\ref{table:flare_list}. We list the cumulative dimming area ($A$); the maximum area growth rate ($\dot{A}$); the total unsigned magnetic flux ($\phi$); the total positive magnetic flux ($\phi_+$); the total negative magnetic flux ($|\phi_-|$); the mean unsigned magnetic flux density ($B_\text{us}$); the maximum brightness drop in the dimming area ($I_{\text{drop}}$); and the maximum brightness drop rate ($\dot{I}_{\text{drop}}$). $I_{\text{drop}}$ and $\dot{I}_{\text{drop}}$ are given in base-difference units.}
    \end{sidewaystable}

    \clearpage
    \twocolumn
    \section{Superposed epoch analysis for the main dimming parameters}\label{Appendix_epoch}
\begin{figure}[th] 
        	\centering
            \resizebox{\hsize}{!}{\includegraphics{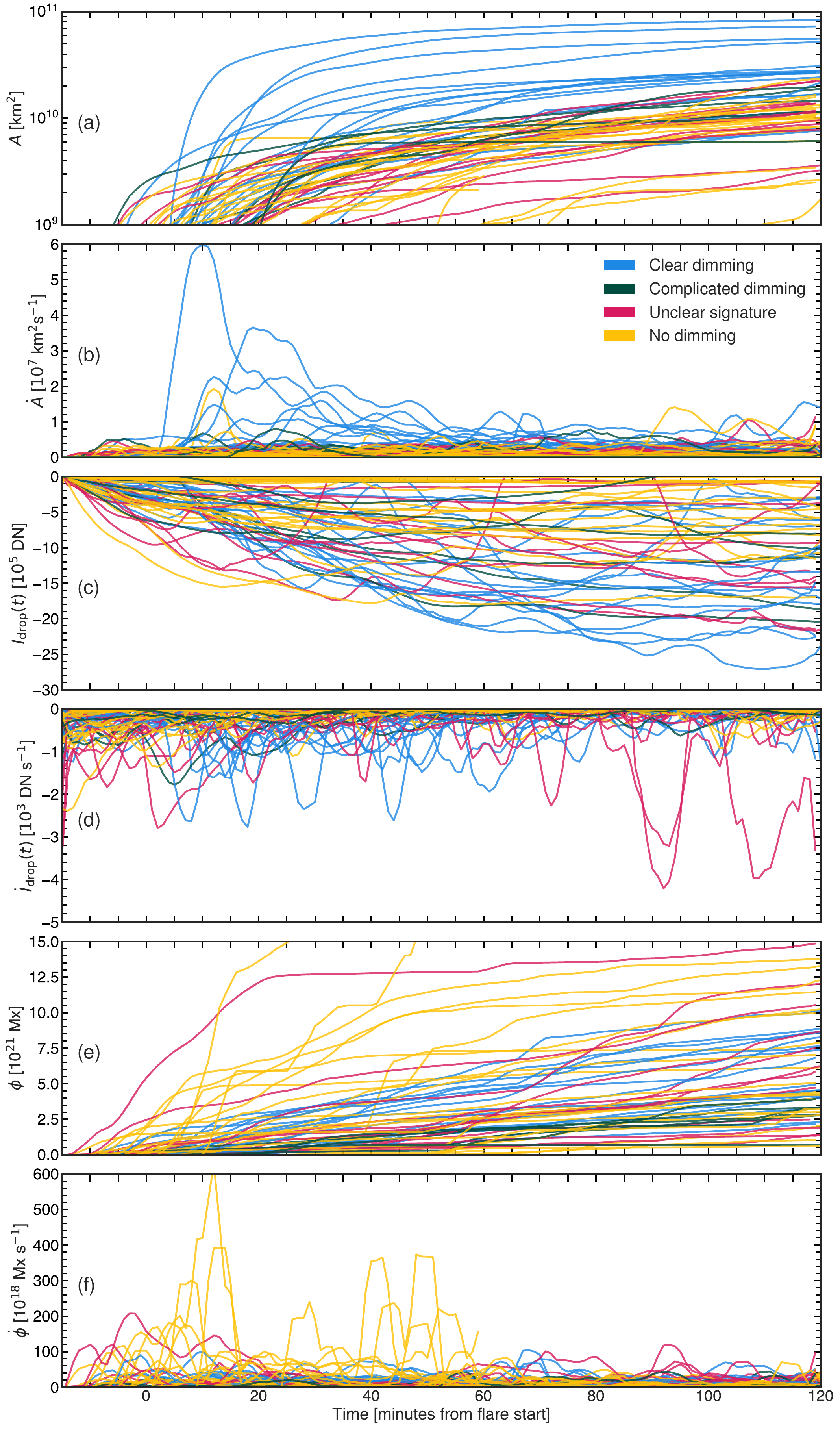}}
        	\caption{Superposed epoch analysis of selected dimming parameters for all flares $\geq$M1.0 class associated with AR~13664 from 2024 May 1 to May 15. Events are colour-coded according to their visual characterisation as dimmings: blue for events with clear dimming signatures, green for dimming events that are detectable but difficult to interpret, pink for events with ambiguous brightness decreases, and yellow for events with no visible dimming. The panels show: (a) Dimming area $A(t)$, (b) dimming area growth rate $\dot{A}(t)$, (c) brightness drop $I_\text{drop}(t)$, (d) brightness drop rate $\dot{I}_\text{drop}(t)$, (e) total unsigned magnetic flux $\phi(t)$, and (f) total unsigned magnetic flux rate $\dot{\phi}(t)$.}
            \label{fig:app:epoch_analysis}
        \end{figure}

    The superposed epoch analysis described in Sect.~\ref{sec:data:selection} was performed for every characteristic dimming parameter. Figure~\ref{fig:app:epoch_analysis} shows the evolution of each parameter with different detections colour-coded based on their visual identification from clear dimming to no dimming. Clear dimmings, indicated in blue, have a characteristic shape only distinguishable in the dimming area $A(t)$ and area growth rate $\dot{A}(t)$.
    
    \section{Alternative data reduction for faulty and/or gap events}\label{Appendix_alternative}
	On May 8, between 16:36~UT and 24:00~UT significant issues are found in SDO data. For SDO/HMI data there is a gap during that period, and for SDO/AIA data some frames appear blurred, rotated and/or off-centre. To account for these issues an alternative data reduction pipeline to that presented in Sect.~\ref{sec:data:reduction} was used.

    Firstly, due to an insufficient amount of AIA images with an exposure time between 1.8 and 3.0~s, frames with larger exposure times were also incorporated. Secondly, blurred frames were manually removed from the analysis. Thirdly, since \texttt{aia\_prep.pro} has trouble processing this data it was omitted and exposure normalisation was explicitly performed instead. To account for the rotation and misalignment, the IDL routine \texttt{rot\_map.pro} was applied. The rebinning and differential rotation were performed analogously to the standard data case. 

    To extract the magnetic properties, the last available HMI image was used, which was taken at 16:36~UT. The image was subsequently processed using \texttt{hmi\_prep.pro} similarly to the rest of the HMI data. Separate HMI maps were created for each flare event, for which the HMI image was rotated to the individual reference times using \texttt{drot\_map.pro}. For the latest event at 21:12~UT, this involved a 4.5~hr shift.

    Importantly, due to the data scarcity, the base image in the affected events was extracted from a single image and not as the combination of three images as described in Sect.~\ref{sec:methods:detection}.

    \section{Correlation coefficients for the May 2024 events dimmings}\label{Appendix_correlations}

        \begin{figure*}[h!]
        	\centering
            \resizebox{\hsize}{!}{\includegraphics{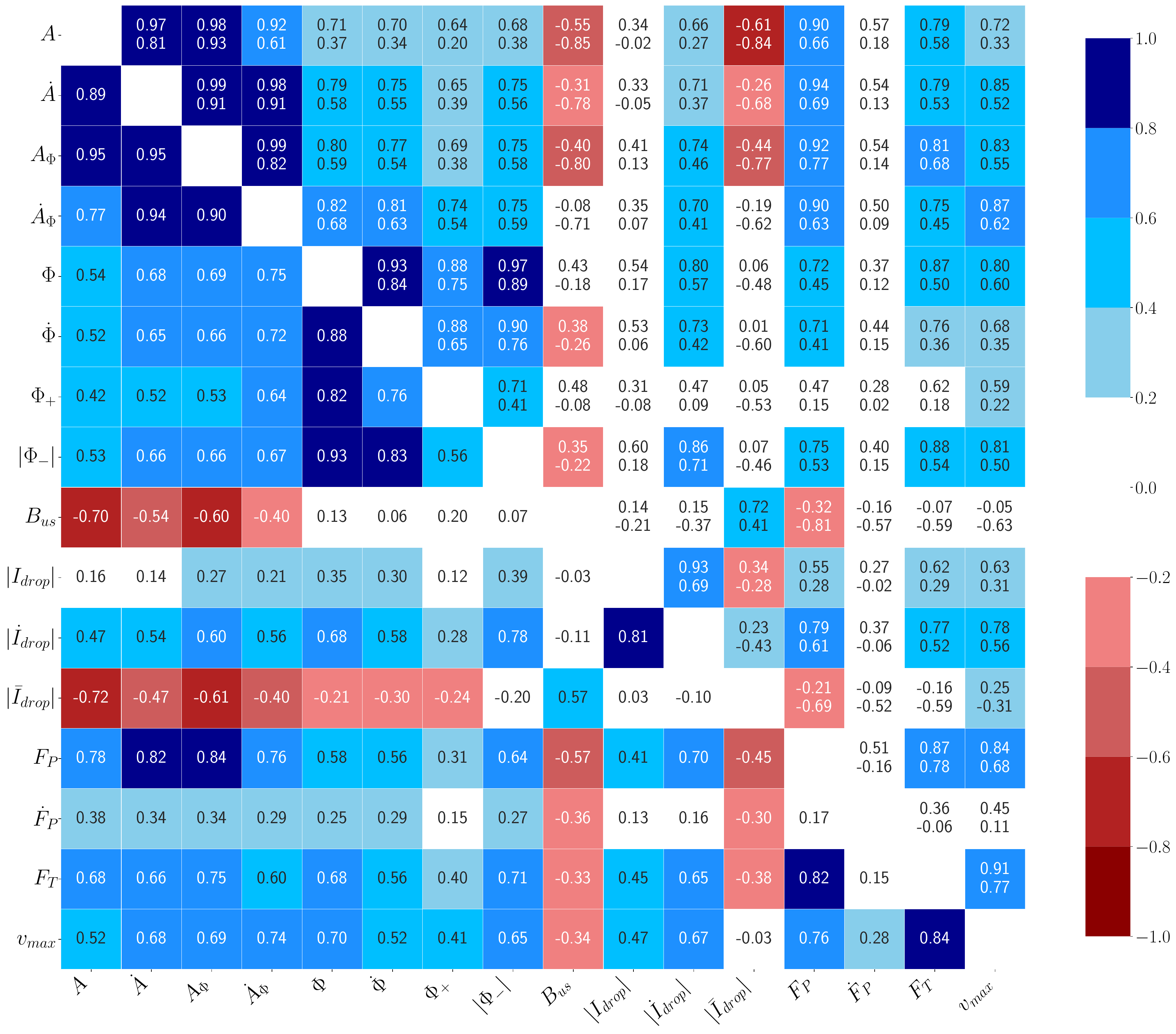}}
        	\caption{Correlation matrix showing the correlation coefficients $c=\bar{c}\pm\Delta c$ of all possible pairs of dimming parameters, flare parameters and CME maximum velocity $v_\text{max}$. The entries below the main diagonal show the mean correlation coefficient $\bar{c}$, while the entries above the main diagonal show two values: $\bar{c}-\Delta c$ (lower value) and $\bar{c}+\Delta c$ (upper value). The colour of the cells indicate the strength of the correlation (blue) or anticorrelation (red). The colours for the cells below the main diagonal are given by $\bar{c}$, while the colours for the cells above the diagonal are given by $\bar{c}-\Delta c$.}
            \label{fig:app:correlation}
        \end{figure*}
        
        The correlation coefficients of characteristic dimming parameters, flare parameters ($F_P$, $F_T$, and $\dot{F}_P$), and maximum CME velocity $v_\text{max}$ are detailed in Fig.~\ref{fig:app:correlation}. 
        
\begin{figure*}[t]
            \centering
            \begin{subfigure}{0.4\textwidth}  
                \centering
                \includegraphics[width=\textwidth]{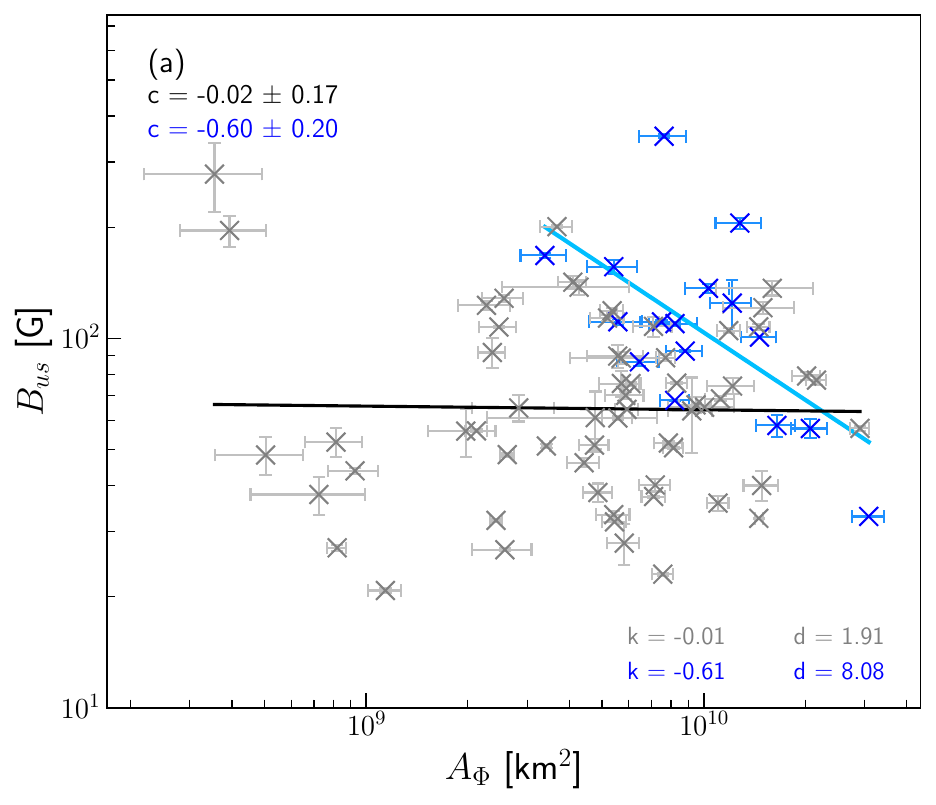}
            \end{subfigure}
            \begin{subfigure}{0.4\textwidth}
                \centering
                \includegraphics[width=\textwidth]{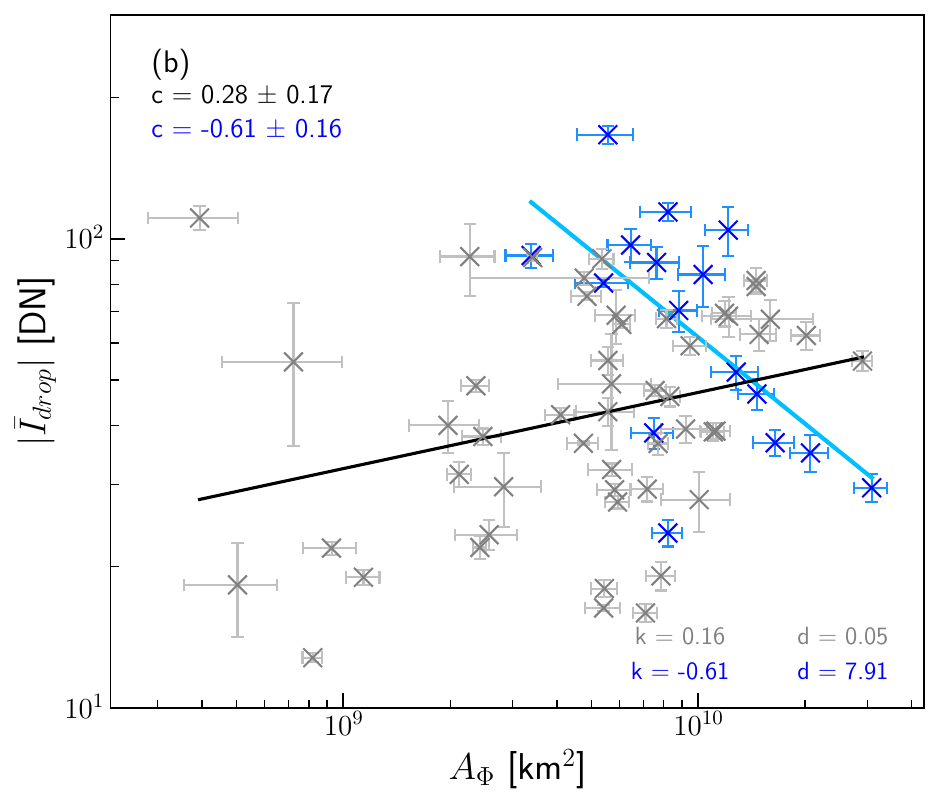}
            \end{subfigure}
            \caption{Correlation plots of (a) the magnetic flux density $B_\text{us}$ and (b) the mean brightness drop $\bar{I}_\text{drop}$ against the magnetic dimming area $A_\phi$. Blue crosses represent dimmings from the May 2024 events, while grey crosses correspond to dimming events adapted from \citet{Dissauer2018b}. The black (blue) regression line is fitted exclusively to the grey (blue) data points.}
            \label{fig:app:anticorrelation}
        \end{figure*}

        Figure~\ref{fig:app:anticorrelation} shows (a) the mean unsigned magnetic flux density $B_\text{us}$ and (b) the mean brightness drop $\bar{I}_\text{drop}$ against the magnetic dimming area $A_\phi$. Both parameters are anticorrelated in the May 2024 events, while the events studied by \citet{Dissauer2018b} show very weak ($\bar{I}_\text{drop}$) to no correlation ($B_\text{us}$). This correlation arises as a consequence of comparing dimmings within the same active region. While the negative trend is visible for the May 2024 events dimmings, the data points do not stand out from the data in \citet{Dissauer2018b} and thus do not indicate any further correlation in the general dimming case.
        
\begin{figure*}[t]
            \centering
            \resizebox{0.7\hsize}{!}{\includegraphics{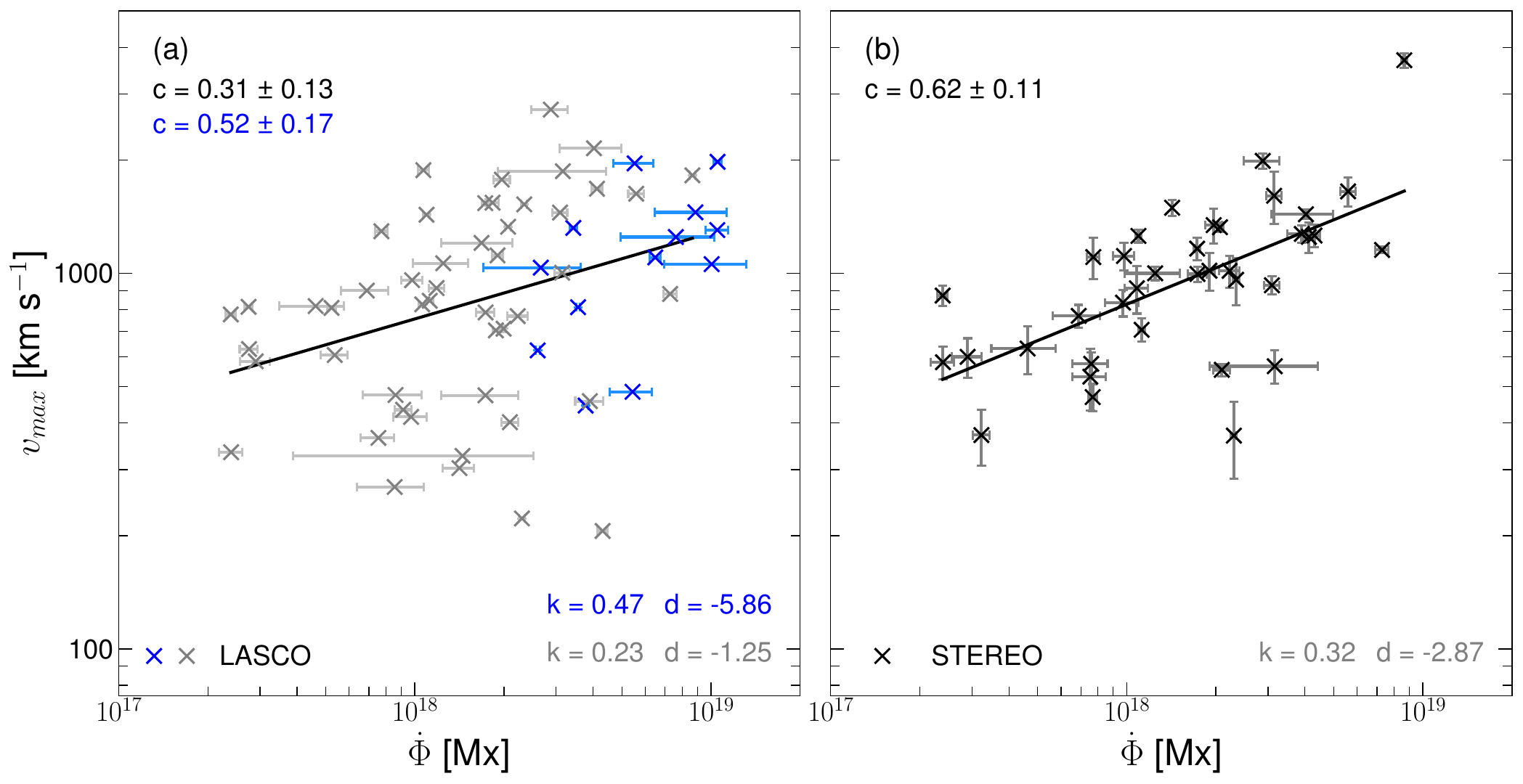}}
            \caption{Correlation plots of the maximum CME velocity $v_\text{max}$ against the dimming magnetic flux rate $\dot{\phi}$. Blue crosses represent dimmings from the May 2024 events, while grey and black crosses correspond to dimming events from \citet{Dissauer2019}. $v_\text{max}$ is extracted from (a) SOHO/LASCO measurements  and (b) STEREO measurements. The black regression lines are fitted exclusively to the grey or black data points.}
            \label{fig:app:correlation_vmax_dphi}
         \end{figure*}

\begin{figure*}[t]
            \centering
            \resizebox{0.7\hsize}{!}{\includegraphics{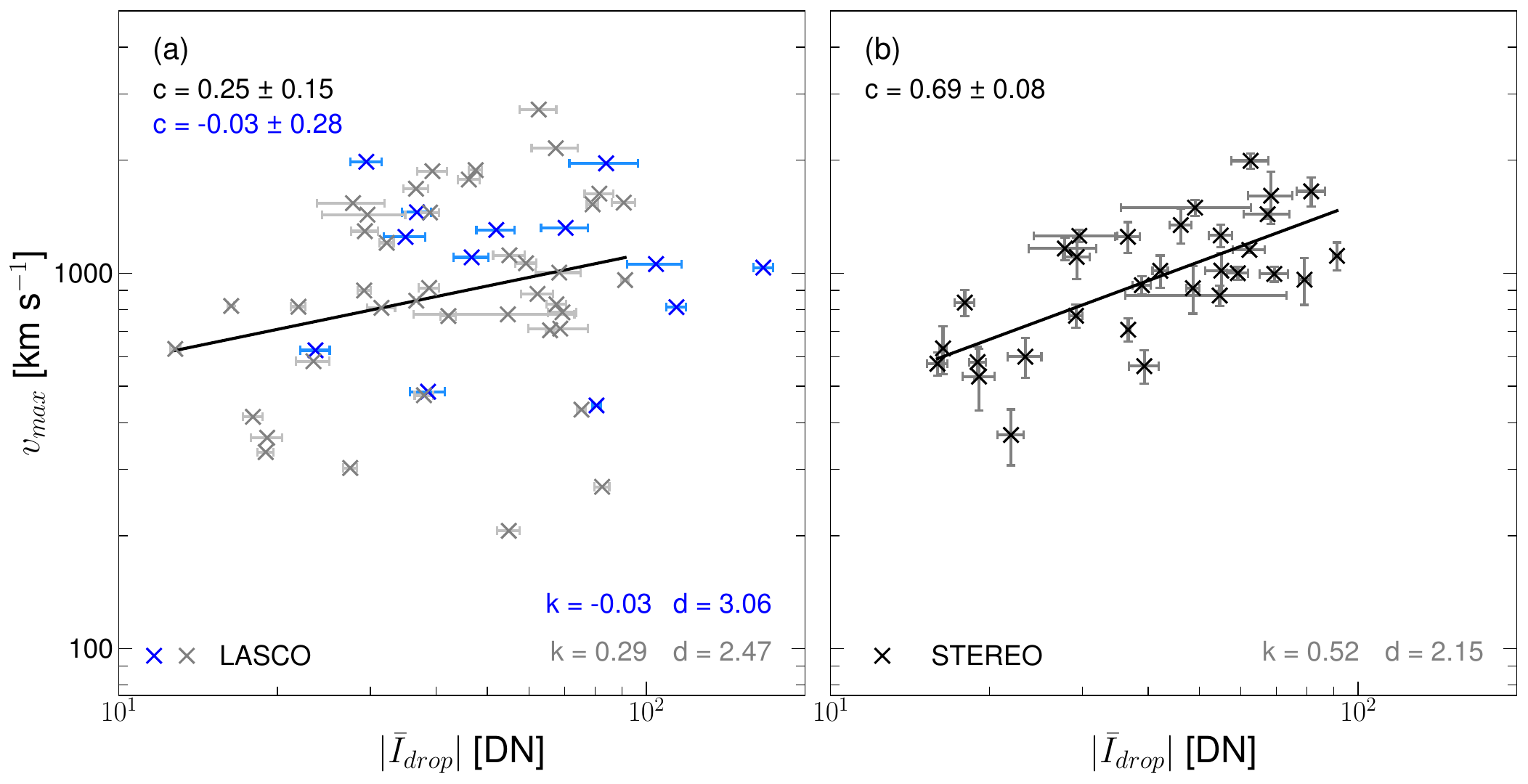}}
            \caption{Same as Fig.~\ref{fig:app:correlation_vmax_dphi} but for the maximum CME velocity $v_\text{max}$ against the dimming mean brightness drop $|\bar{I}_\text{drop}|$.}
            \label{fig:app:correlation_vmax_Imean}
         \end{figure*}

        Figures~\ref{fig:app:correlation_vmax_dphi} and \ref{fig:app:correlation_vmax_Imean} illustrate the difference in correlation between CME maximum velocity $v_\text{max}$ and characteristic dimming parameters ($\dot{\phi}$ and $\bar{I}_\text{drop}$, respectively) when the velocity is extracted using only coronagraph data compared to using also EUV imagers. Panel (a) in both figures contain $v_\text{max}$ calculated from heigh-time profiles observed with LASCO, while in panel (c) the black data points were calculated using STEREO observations by \citet{Dissauer2019}. The STEREO-based velocities show a narrower distribution against $\dot{\phi}$ and $\bar{I}_\text{drop}$, and have clear correlations compared to the LASCO-based observations.

\end{appendix}

\end{document}